\documentclass[usenatbib]{mn2e}
\usepackage{graphicx}

\title[The gas kinematics in Mrk 533: a gaseous outflow]
{The gas kinematics in Mrk 533 nucleus and circumnuclear region: a
gaseous outflow}

\author[A. A. Smirnova et al.]{A. A. Smirnova$^{1}$\thanks{E-mail:
alexiya@sao.ru}, N. Gavrilovi\' c$^{2,3}$, A. V. Moiseev$^1$, L. \v
C. Popovi\'c$^{2}$, V. L. Afanasiev$^{1}$,
\newauthor P. Jovanovi\' c$^{2}$, M. Da\v ci\'c$^{2}$\\
$^{1}$Special Astrophysical Observatory, Nizhnij Arkhyz 369169, Russia\\
$^{2}$Astronomical Observatory, Volgina 7, Belgrade, 11160, Serbia\\
$^{3}$Universit\'e de Lyon, Lyon, F-69000, France ; Universit\'e
Lyon~1,
Villeurbanne, F-69622, France  Centre de Recherche Astronomique de
Lyon, \\ Observatoire de Lyon, 9 avenue Charles Andr\'e, Saint-Genis
Laval cedex, F-69561, France ; CNRS, UMR 5574 ; \\
Ecole Normale
Sup\'erieure de Lyon, Lyon, France}

\begin{document}



\maketitle

\label{firstpage}

\begin{abstract}
We present an analysis of  3D spectra of Mrk 533, observed with the
integral-field spectrograph MPFS and using the Fabry Perot Interferometer
(FPI)  of the SAO RAS 6-m telescope.  We found emissions of gas from the
active Sy 2 nucleus in the centre and also from the HII regions in a spiral
structure and a circumnuclear region. The gas kinematics shows regular
non-circular motions in the wide range of  galactocentric distances from 500
pc up to 15 kpc. The maps of inward and outward radial motions of the ionized
gas was constructed. We found that the narrow line region (NLR) is composed of
at least two (probably three) kinematically separated regions. We detect a
stratification in the NLR of Mrk 533 with the outflow velocity ranging from
20-50 km\,s$^{-1}$ to 600-700 km\,s$^{-1}$, respectively, on the radial
distances of $\sim2.5$ and $\sim1.5$ kpc. The maximal outflow velocity comes
from the nucleus and corresponds to the position of the observed radio
structure, which is assumed to be created in an approaching jet. We suggest
that these ionized gas outflows are triggered by the radio jet intrusion in an
ambient medium.
\end{abstract}


\section{Introduction}

Theoretical studies have shown that the interactions of galaxies can
bring gas from the disc towards the nuclear regions, and that they can also
produce a burst  of star formation  \citep{perez}. Therefore, the
detailed investigation of the emission of gas kinematics in nearby
Seyfert galaxies is very important. The panoramic (3D)
spectroscopy is a powerful tool for this goal. Here we present a
 spectroscopic study of Mrk 533, an interacting Sy 2 galaxy.
Mrk 533 (also denoted as NGC 7674, Arp 182, UGC 12608, H96a) is the brightest
member of the Hickson 96 compact group of galaxies \citep{Hick82}. It is the
only spiral (Sbc pec)  galaxy among other tidally interacting galaxies in the
group \citep{Ver97}. The kinematics of Mrk 533 was observed twice  with
scanning Fabry-Perot systems \citep{afak96,amram03} and once using the long
slit-spectra technique \citep{Ver97}. \citet{afak96} mapped the velocity field
in the H$\alpha$ line observed with the 6-m telescope using FPI. The spectral
resolution was $\delta\lambda\approx1.0$ \AA\, and the free (interfringe)
spectral range was $\Delta\lambda\approx13$ \AA. They also used an
integral-field data and reported a complex structure of emission lines in the
central region within $r<5$ arcsec, where the H$\alpha$ has a broad and narrow
component. Moreover, the narrow component could be also divided into two
Gaussians, where one of them corresponds to the normal rotation and  second
seems to be connected with the nuclear radio structure. \citet{afak96} found
non-circular motions that are probably caused by the circumnuclear bar.
\citet{amram03} also observed Mrk 533 and its environment with 3.6-m ESO
Telescope using the FPI system ($\delta\lambda\approx0.7$ \AA\, and
$\Delta\lambda\approx8.5$ \AA). They found a significant disagreement in shape
of radial velocity distribution between receding and approaching sides of the
galactic discs at  distances $r>20$ arcsec. A possible shift of the rotation
centre around $~2$ arcsec with respect to the optical centre was also noted.
Both features are probably caused by interactions with companions, especially
with HCG96c.

 Mrk 533 is a Sy 2 galaxy where the broad H$\alpha$ and H$\beta$
lines are observed in polarized flux \citep{Mill90}, indicating
the presence of a hidden Broad Line Region \citep{Tran95}. In
addition, the Narrow Line Region of Mrk 533 is extended, as shown
in the [OIII]$\lambda\lambda$4959,5007 image observed with HST
\citep{Schm03}. \citet{Unger88} found that  radio maps reveal the
presence of a triple radio source with a total angular extent of
about 0.7 arcsec. This provides evidence that the radio emission is
powered by  ejection. In the plane of the sky, the
ejection axis is found to be roughly perpendicular to the galactic
rotation axis \citep{Unger88}. Recently,  VLBI continuum and HI
absorption observations of the central part with 100 mas
resolution, showed six continuum structures extending over 1.4
arcsec (742 pc), with a total flux density of 138 mJy
\citep{Mom03}. \citet{Mom03} also suggested that the overall
S-shaped radio pattern could be the result of a interstellar
medium diverting the out-coming jets from the central AGN. In the same time,
they could not  role out the possibility of a black hole merger
that could result in a similar structural pattern.

The galaxy has noticeable blue-shifted wings on almost all of the
emission lines (except [SII]$\lambda\lambda$6716,6730\AA) as
originally suggested by \citet{Afan80} and later confirmed by
other authors (e.g. \citet{Shuder81}, \citet{Laur88},
\citet{Rob90}, \citet {Vei91}, etc.). Moreover, the observed UV
doublet O VI$\lambda\lambda$1032,1038 line profile shows a blue
wing that indicates an outflow in the UV emission gas
\citep{Shas04}. Also, the absorption lines at -300 km\,s$^{-1}$
and -800 km\,s$^{-1}$ have been observed in the O VI doublet
\citep{Shas04}. The absorption in the O VI and a blue asymmetry in
the optical lines might be caused by an accelerated wind from the
central source.

Here we present an analysis of Mrk 533 spectra observed with 6-m
telescope of the SAO RAS using the integral-field spectrograph MPFS and
Fabry-Perot Interferometer (FPI). The aims of the paper are: (i)
to investigate the emission line structures of Mrk 533, focusing on the
central part in order to map gaseous outflows indicated in
the previous radio and UV observations; (ii) to study the gas
kinematics in the circumnuclear region. Also, here we will discuss
the kinematics of a gaseous disc as a whole which has a non-circular
component of motion  in order to find the link between the
host galaxy and central part kinematics.

In this paper we adopt the value of 116
Mpc given by \citet{Mom03}, to be the Mrk 533 distance, using the scale 1  arcsec=563 pc.

 The paper is structured as follows: In Section~\ref{obs} we present our
observations, in Section~\ref{kin} we analyze the gas kinematics
in Mrk 533, in Section~\ref{nlr} we investigate in more detail the
complex NLR in this galaxy, in Section~\ref{rad} we discuss
the possible link between the radio jet and optical one, and finally in Section~\ref{concl} we give our conclusions.

\begin{center}
\begin{table*}
\caption{The spectroscopic observations of Mrk 533.}
\label{speclog}
\begin{tabular}{llllll}
\hline
Date & T$_{exp}$ & Sp. Range & Sp. Res. & Seeing &  Instrument \\
& (sec) & (\AA) & (\AA)  &  (arcsec) & \\
\hline
1998 Aug. 30 &  $24 \times180$ &  H$_{\alpha}$  &  3.0  & 1.5  &  FPI     \\
2002 Aug. 13 &  3600     &    4500 - 7100   &  7.5  & 1.5  &  MPFS    \\
2005 Sep. 28 &  6000     &    4350 - 5900   &  4.2  & 1.3  &  MPFS    \\
\hline
\end{tabular}
\end{table*}
\end{center}

\begin{figure}
\includegraphics[width=7cm]{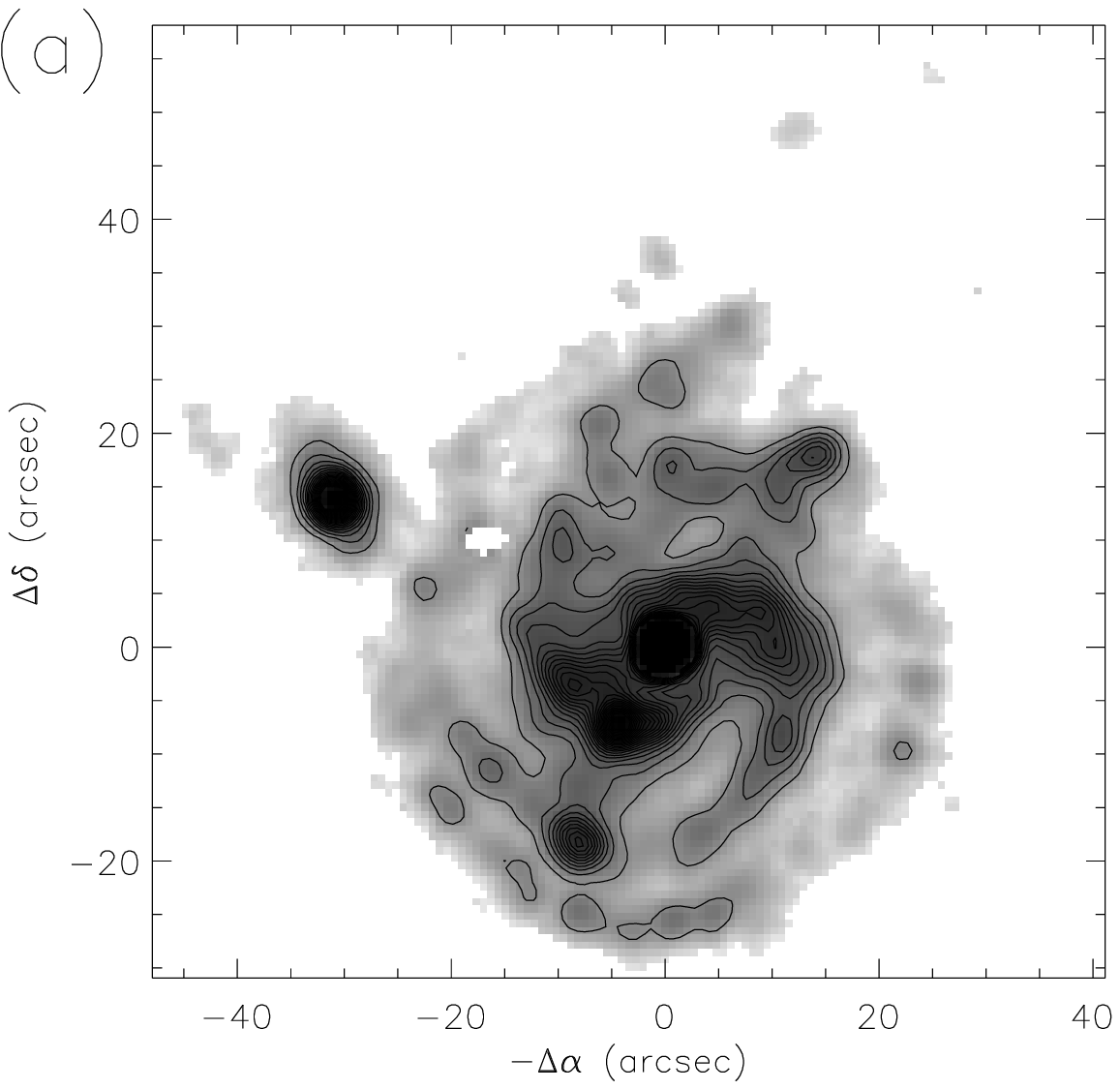}
\includegraphics[width=7cm]{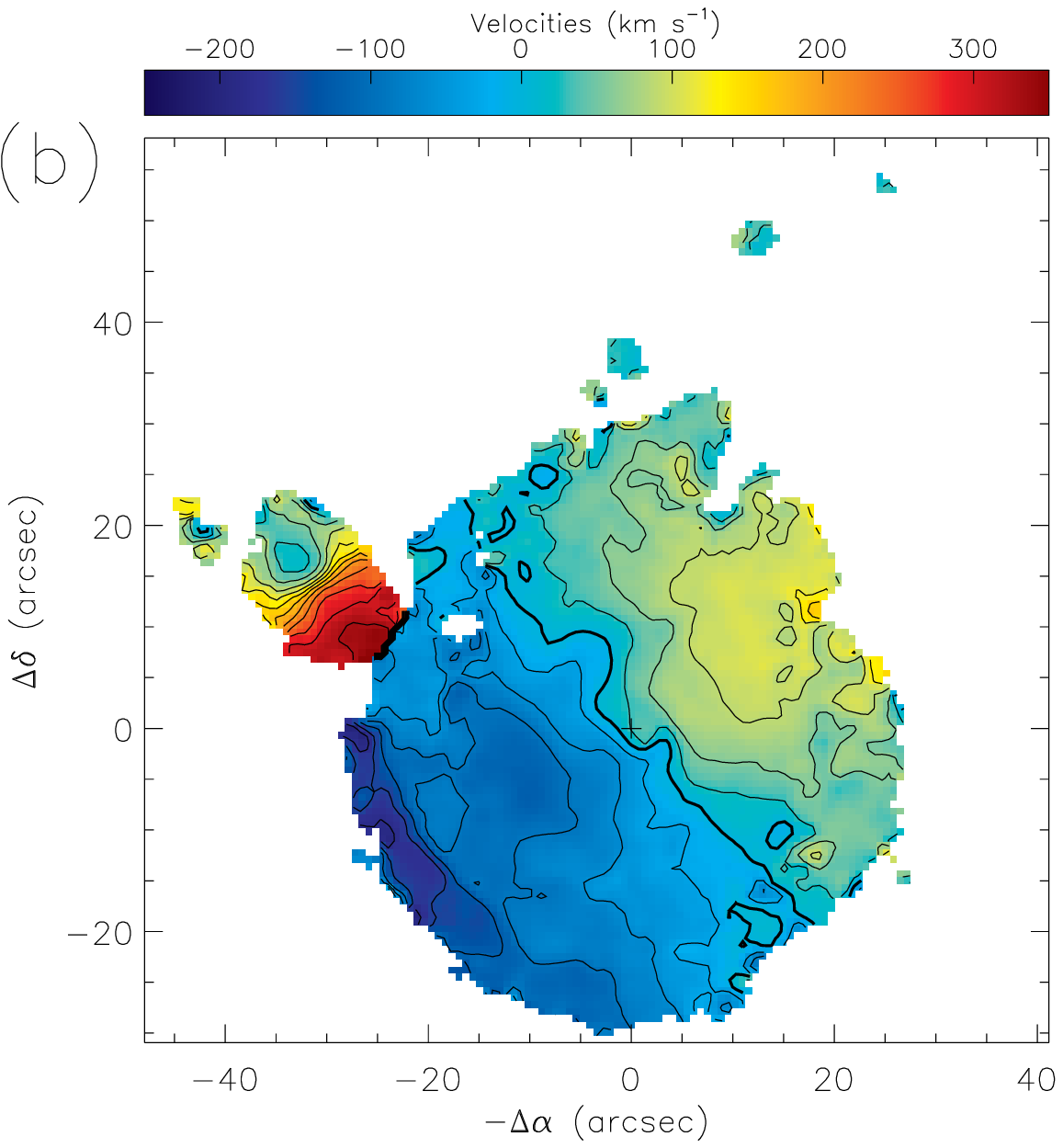}
\includegraphics[width=7cm]{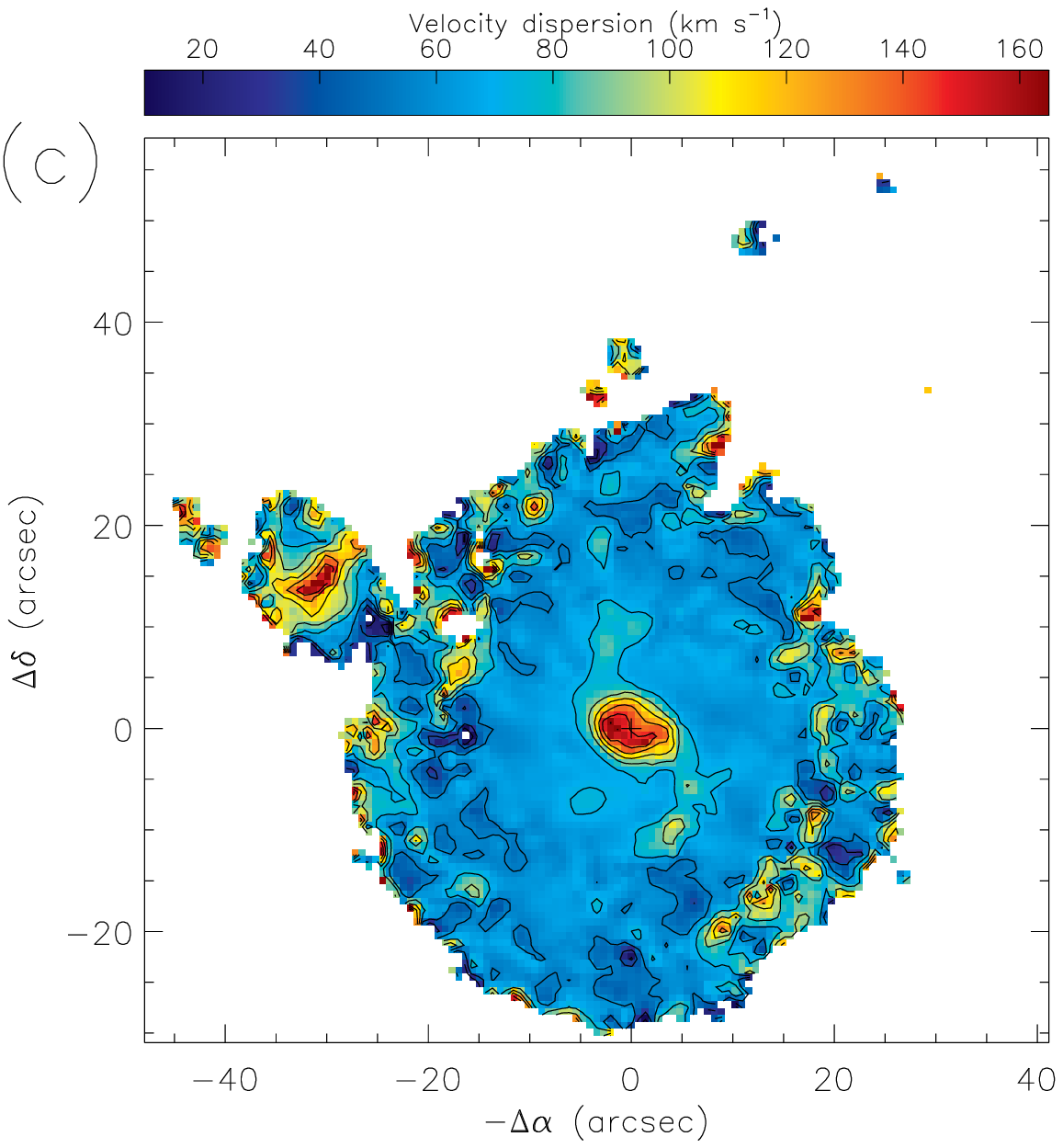}
\protect\caption{ Large-scale emission-line intensity distribution
in log-scale (a), ionized-gas line-of-sight velocity field minus
the systemic velocity (b) and velocity dispersion  (c) for Mrk~533
obtained from FPI data for the H$\alpha$ line.  The cross
corresponds to the continuum centre. The thick contour on (b)
marks the zero velocity.} \label{fig_01}
\end{figure}

\section{Observations and data reduction}\label{obs}

The panoramic (or 3D) spectroscopy provides spectra of extended object
simultaneously for each spatial element in a two-dimensional field of view
(FOV). We used  two different types of 3D spectrographs: the integral-field
spectrograph MPFS and scanning FPI. The first device provides a detailed
spectral information in the spectral range about  several thousand \AA\, what
is important for the goals of the spectrophotometry and for analyzing  the
broad line profiles. However, the FOV of MPFS is small (less than 16 arcsec),
it can cover only a region around the active nucleus of Mrk 533. In contrast,
the scanning FPI had a very large FOV (a few arcmin) with a very narrow
spectral region containing only a spectral line. Using this device we are able
to  investigate the long-scale kinematics of the ionized gas in the whole
stellar disc of Mrk 533, but only in the narrow component of the  H$\alpha$
emission line.

\subsection[]{Integral-Field Spectrograph MPFS}

Mrk 533 was observed in August 2002 and September 2005 using the MultiPupil
Fiber Spectrograph (MPFS), with the integral-field unit mounted at the primary
focus of the 6-m telescope \citep*{Afa01}. The MPFS takes simultaneous spectra
of 240 spatial elements (constructed in the shape of square lenses) that form
on the sky an array of $16\times15$ elements (in 2002), or of 256 spatial
elements ($16\times16$ in 2005). The angular size was 1 arcsec per element in
2002, and 0.75  arcsec per element in 2005. The detectors were a CCD TK1024
($1024 \times 1024$ px) in 2002 and EEV42-40 ($2048 \times 2048$ px) in 2005.
A description of the MPFS is available at SAO RAS web page {\it
http://www.sao.ru/hq/lsfvo/devices.html}.

During the first run, in August 2002, for spectrophotometry purposes, the
galaxy was observed at a low resolution, while during the second run it was
subject to a kinematics investigation of the NLR, and was observed at a higher
resolution. The log of MPFS observations is given in Table~\ref{speclog}.

The data were reduced using the software developed at the SAO
RAS by V.L. Afanasiev and A.V. Moiseev and running in the IDL
environment. The primary reduction included bias subtraction,
flat-fielding, cosmic-ray hits removal, extraction of individual
spectra from the CCD frames, and their wavelength calibration
using a spectrum of a He-Ne-Ar lamp. Subsequently, we subtracted
the night-sky spectrum from the galaxy. The spectra of
the spectrophotometry standard stars were used to convert counts into
absolute fluxes.

\subsection{Scanning Fabry-Perot Interferometer}

The galaxy was observed in August 1998 with the scanning FPI. The Queensgate
interferometer ET-50 was used in the 235th interference order (for the
H$\alpha$ line). The free spectral range between neighboring orders
(interfringe) was about 30\,\AA. For monochromatization a narrow-band filter
($FWHM=31$\AA) centred  on the spectral region containing the redshifted
H$\alpha$ was used. During the observations we successively took 24
interferometric images of the object with different gaps between the FPI
plates. The spectral channels were of $\delta \lambda\approx1.2 $ \AA\ width,
($\sim 56\ $ km\,s$^{-1}$), the spectral resolution (the width of instrumental
contours) was $FWHM\approx 3.0$~\AA\, ($\sim140$ km\,s$^{-1}$). The detector
was a CCD TK1024 ($1024\times1024$ pixels) operated by $2\times2$ binning for
reading-out time economy. Therefore each spectral channel had a $512\times512$
pixels format with the scale 0.68 arcsec  per pixel. The log of the FPI
observations is presented in Table~\ref{speclog}.

To reduce the interferometric observations we used a special
software \citep{moi02}, running in the IDL
environment. After the primary reduction (bias, flat-field, cosmic
hits), we removed the night-sky spectrum, converted the data to
a wavelength scale, and prepared them as a `data cube'. The
spatial resolution after data reduction smoothing was about 2.5 arcsec.
The velocity fields of the ionized gas, and images in the
H$\alpha$ emission line were mapped using the Gaussian fitting of the
emission line profiles. Moreover, we created an image of the
galaxy in the  continuum close to the emission line.

\begin{figure}
\includegraphics[width=8.8cm]{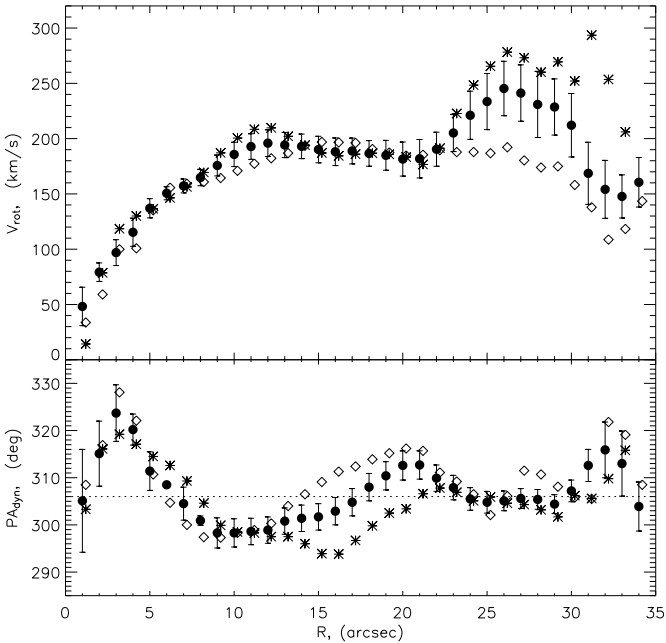}
\protect\caption{Results of the tilted-ring analysis of the
H$\alpha$ velocity field: radial variations of the rotation
velocities ({\it top}) and position angles ({\it bottom}). Full
circles, stars and open diamonds denote the value calculated on
the whole disc, SE (approaching) and NW (receding) sides
respectively. Dotted line corresponds to $PA_0=306^\circ$ adopted
in this paper (see for details Section \ref{circ}).}\label{fig_02}
\end{figure}

\section{The kinematics of Mrk 533}\label{kin}

\subsection{Analysis of FPI observations}

Both  previous observations were performed using the FPI \citep{afak96,amram03}
with a high spectral resolution $\delta\lambda=0.7-1.0$
\AA\, but within a small spectral range ($\Delta\lambda=8.5-13$ \AA).
We performed new observations with the 6-m telescope with a
relatively low spectral resolution ($\delta\lambda\approx3.0$
\AA), but with a larger $\Delta\lambda\approx30$ \AA.
It is important for the study of the emission line profile in the
nucleus, where FWHM of lines reaches $8-10$ \AA. Unfortunately,
this spectral range is insufficient for a detailed analysis of
emission line profiles, with FWHM larger than 30 \AA\, (see
Section~\ref{nlr}). Therefore, with FPI we have considered only
kinematics of the narrow component of the H$\alpha$. For investigation,
the broad component and the blue-wing asymmetry of the [OIII] lines
 we have used MPFS data only.

The H$\alpha$ monochromatic image, obtained with FPI, is shown
in Fig. \ref{fig_01}a. As one can see in Fig. \ref{fig_01}a,
besides the strong central part in the H$\alpha$, there are
structures (which follow the bar and spiral arms) which
show a significant H$\alpha$ emission. The ionized gas velocity
field and velocity dispersion map (corrected on the instrumental
contour) are shown in Figs \ref{fig_01}b,c.

As can be seen in Fig. \ref{fig_01}b, the  rotation  axis is SW-NE
oriented. It is interesting to note that in  the velocity map two
features with  approaching velocities (with respect to the
systematic one) are present. These features will be discussed in
more details in Subsection~\ref{cnlr}. The velocity dispersion map
(Fig. \ref{fig_01}c) shows the arms near the centre  and, as
expected, the maximum velocity dispersion in the central part.

\begin{figure*}
\includegraphics[width=8.5cm]{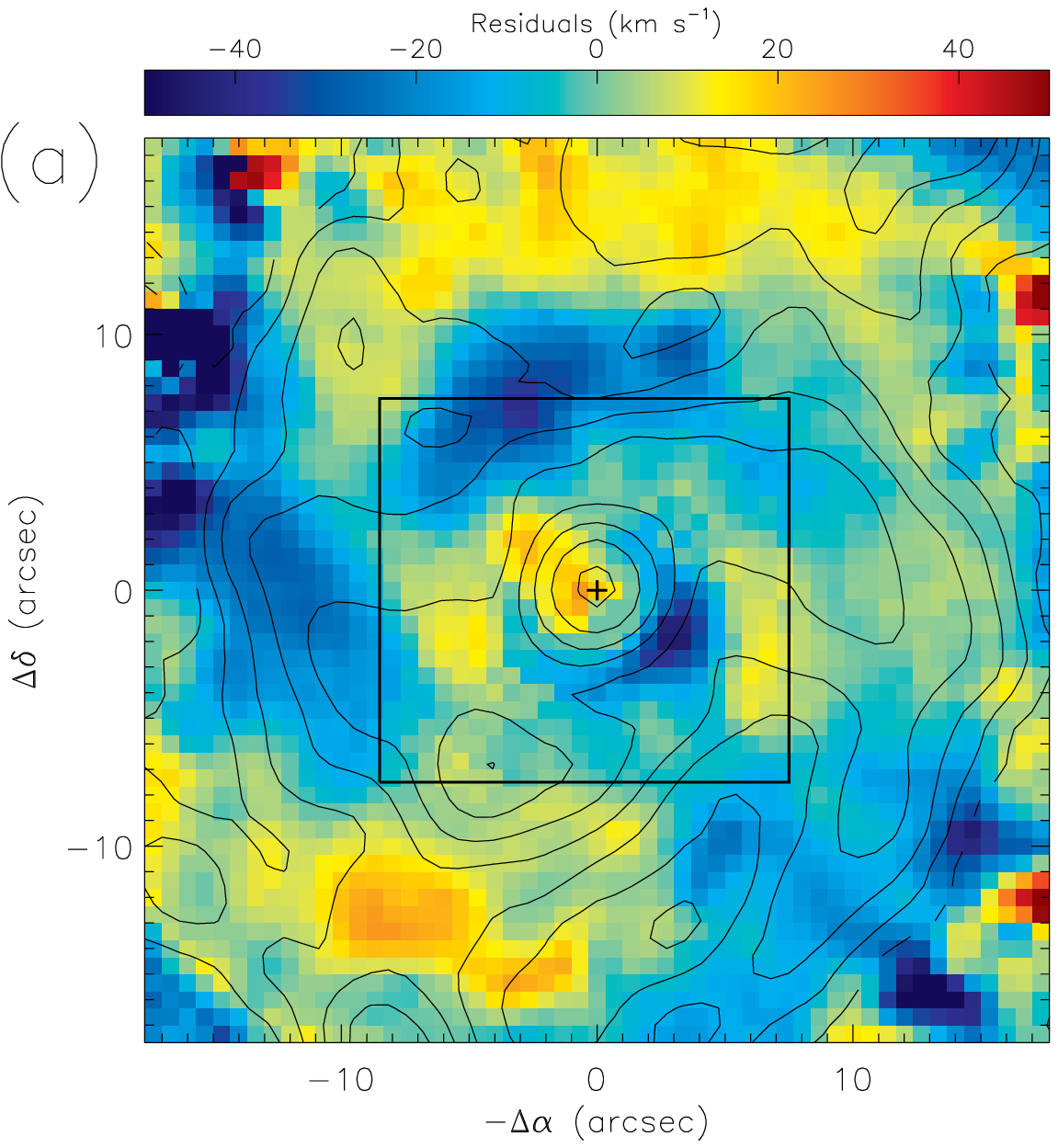}
\includegraphics[width=8.5cm]{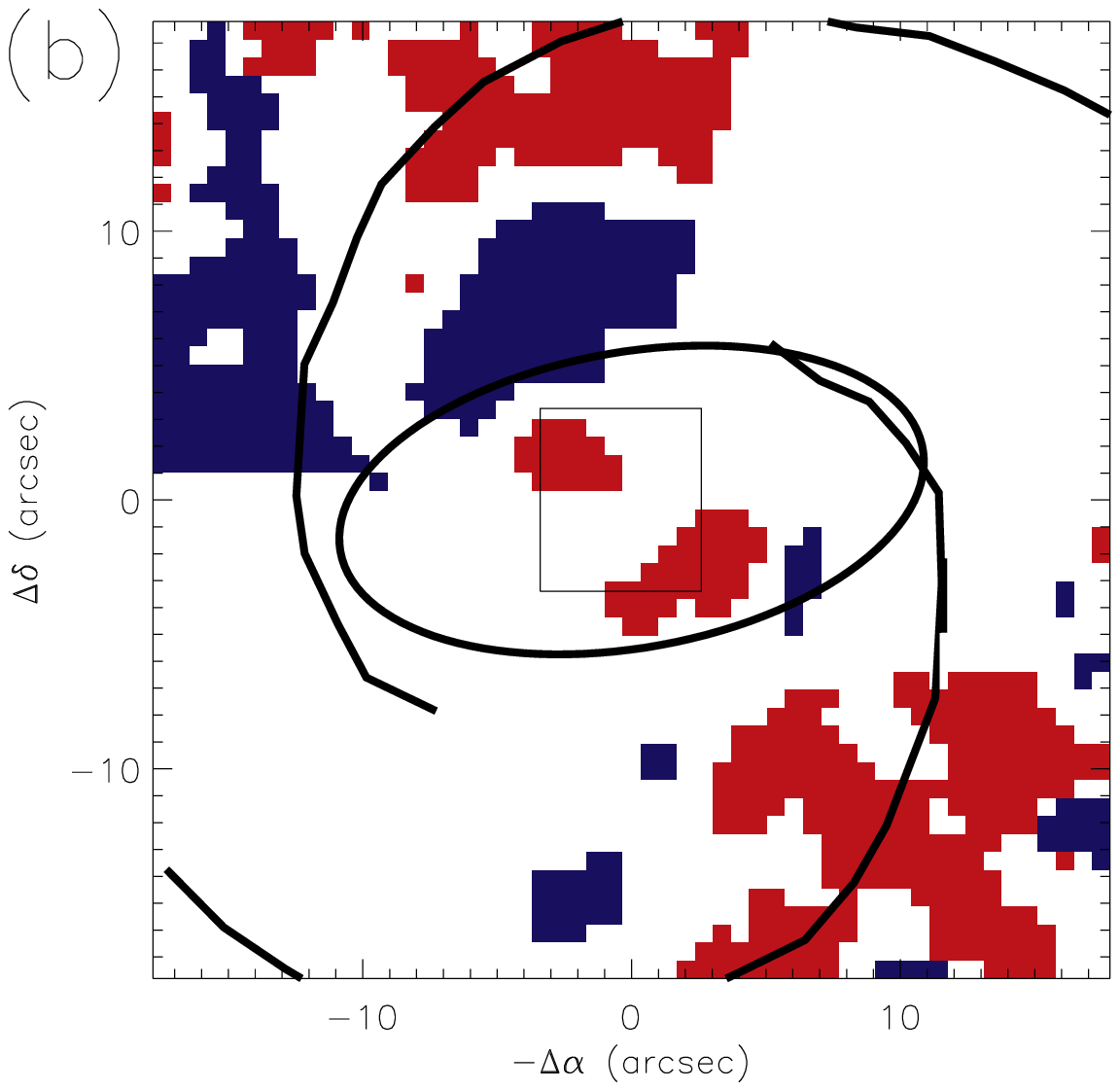}
\protect\caption{Ionized gas kinematics in the central region: (a) residual
velocities after subtraction of the model of pure circular rotation from the
observed velocity field, H$\alpha$ isophote are overlapped, the box
corresponds to MPFS field from Fig. \ref{fig_10n}b; (b) the map of radial
motions: 
the inflow is denote with blue and the outflow is red. The galactic bar and
spiral arms are shown. The central region mapped with MPFS (Figs.
\ref{fig_05}-\ref{fig_09}) is shown as a small box in the
centre.}\label{fig_03}
\end{figure*}

Our maps seem to be  slightly deeper  than the ones given by
\citet{amram03}. For instance, we detected the velocities in
several emission clumps along tidal tail located at the North of
the galaxy. The field-of-view contains also a nearby satellite
HCG96c located to the NE from the Mrk 533. We measured the systemic
velocity of this companion, and obtained $8850$ km\,s$^{-1}$
(z=0.0295). This is in  good agreement with the value given by
\citet{Ver97}, and different around $\sim200$
km\,s$^{-1}$ 
from the results of \citet{amram03}.

\subsection{The model of pure circular rotation}\label{circ}

The velocity field was analyzed using the so called   `tilted-rings' method by
\citet{begeman}\footnote{See also \cite*{mois04} and references therein for
more details.}. We were looking for the parameters  of the galactic gaseous
disc orientation, i.e. a line of node position angle (major axis) $PA_0$ and
the galactic plane inclination  $i_0$. The velocity field was fitted using a
pure circular rotation model by fixing the centre position and systemic
velocity. We excluded from the analysis  the central region ($r<7$ arcsec)
where strong non-circular motions are present (see Section~\ref{nlr}) and the
external points ($r>25$ arcsec) where the velocity field is significantly
asymmetrical, as already noted by \citet{amram03}. The following values were
found from a $\chi^2$-minimization routine of residuals between the observed
and modeled velocity fields: $i_0=(33\pm5)^\circ$, $PA_0=(306\pm4)^\circ$. We
assumed that the position of the kinematic centre corresponds to the velocity
field centre symmetry. The obtained systemic velocity is $V_{sys}=8672\pm3$
km\,s$^{-1}$ (z=0.0289). The kinematic centre of the galaxy is shifted for 1
pixel (or $0.7$ arcsec) to the South from the continuum centre. This centre
offset (lopsidedness) was already mentioned by \citet{amram03} and
\citet{Ver97}, however different authors found different positions of the
centre offset. The observed shift of the kinematic centre might be due to the
interaction with the companion as well as with matter that has a non-circular
motion caused by the bar's gravitational potential.

The orientation parameters are in   accordance with the ones measured by
\citet{Ver97}, $PA=121^\circ$ (i.e. $301^\circ$), $i=31^\circ$, while these
values are  significantly different to the ones reported by \citet{amram03},
$PA=132\pm5^\circ$ and $i=50\pm5^\circ$.

\subsection{Detailed analysis of the H$\alpha$ velocity field}\label{vfield}

In order to study  the nature of the deviations from the normal rotations, the
velocity field was divided into elliptical rings with a $1$ arcsec width
aligned to the accepted $PA_0$ and $i_0$. Using the $\chi^2$ minimization
routine for deviations of the observational points from the model, we
calculated, for each radius, the position angles of the major axis
$PA_{dyn}(r)$ (dynamical position angle) and the mean of rotational velocity
$V_{rot}(r)$. The results of the best-fitting are shown in Fig. \ref{fig_02},
where results  for approaching (asterisk in Fig. \ref{fig_02}) and receding
sides (diamonds in Fig. \ref{fig_02}) are shown. These are calculated
separately. As was noted in \citet{amram03} earlier, we obtained a different
kinematics between these two sides.

The mean (`in the whole disc') rotation curve shows a very unusual
shape at distances larger than $r=20$ arcsec. The maximum is at $r\approx26$ arcsec,
after that the rotational velocity decreases. As it is shown in  Fig. \ref{fig_02}, this maximum
in the mean rotational curve is caused by increase of the
rotation velocity (after $r>20$ arcsec) in the approaching (SE)
side of the disc, while $V_{rot}(r)$ of the receding side is
slightly decreasing in this region. Also, a small $V_{rot}(r)$
peak in the approaching side can be noticed between 10 and 14
arcsec. These features cannot be caused by specific mass
distribution, but rather by an interaction with
other members of the Hickson group, especially at the maximum
$r>20$ arcsec.

\begin{figure*}
{ \includegraphics[width=5.5cm]{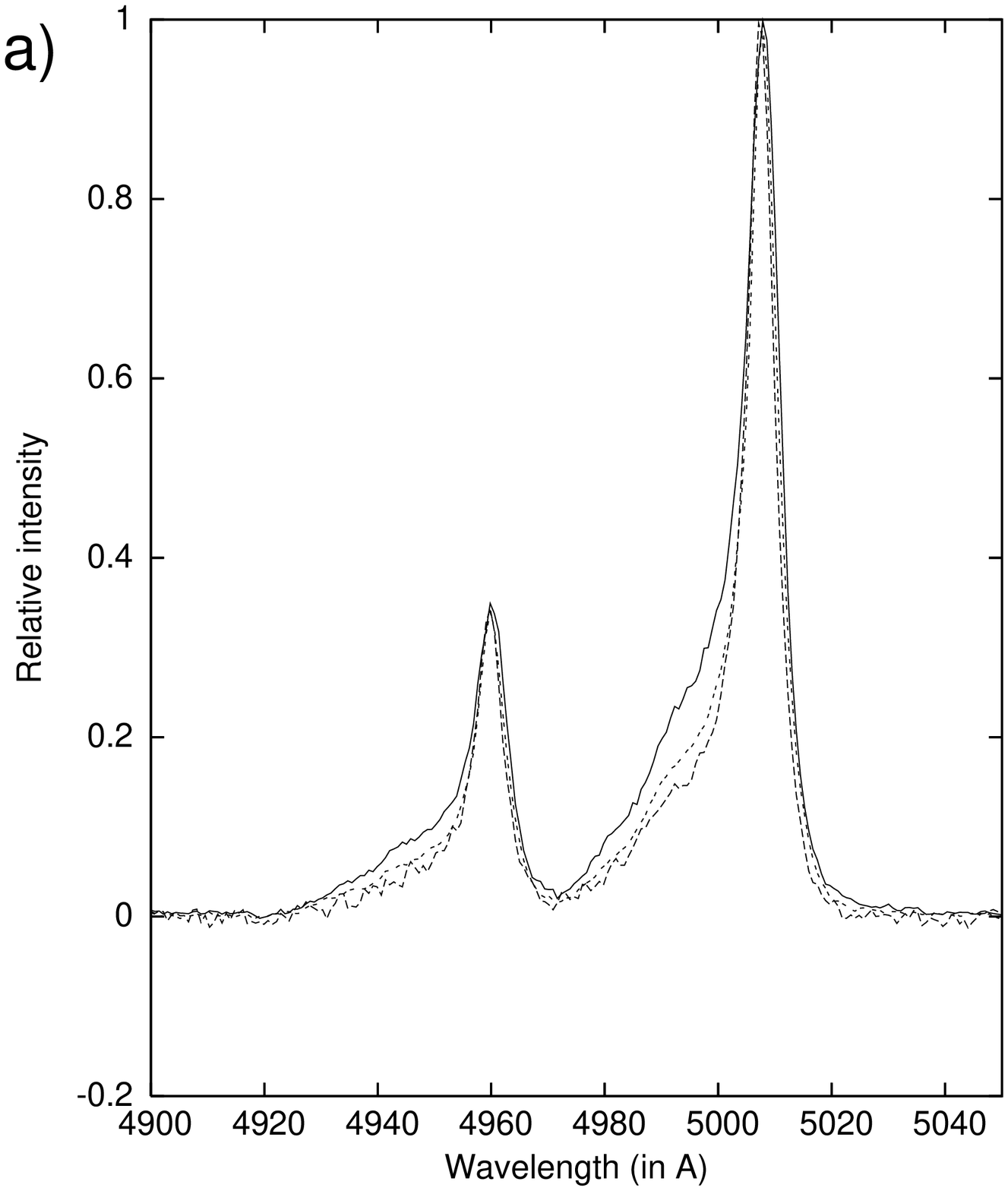}}
{\includegraphics[width=5.5cm]{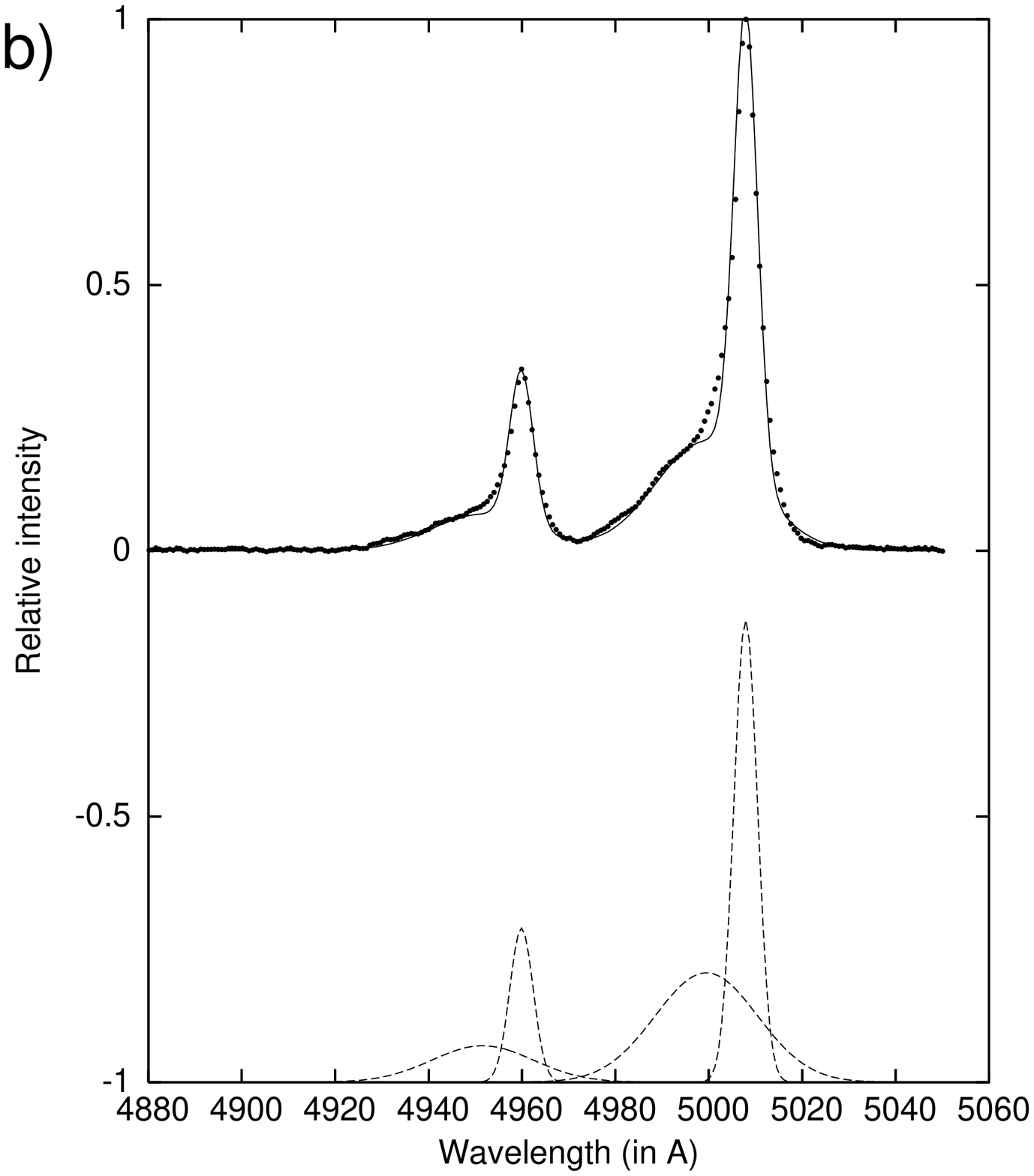}}
{\includegraphics[width=5.5cm]{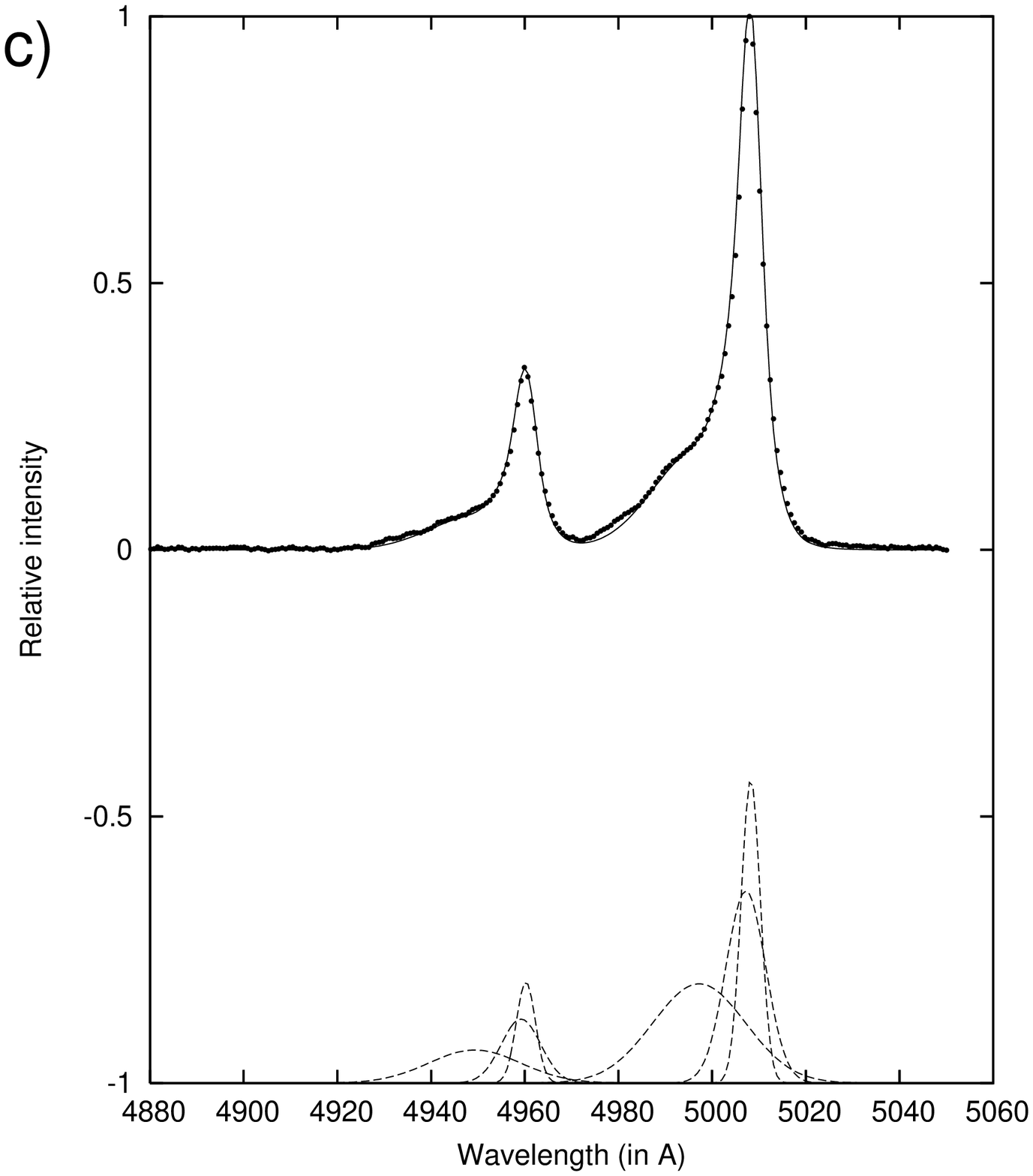}} \caption{Difference in
the blue part of the [OIII] lines across the central (nucleus)
part of Mrk 533
 (a).  Decomposition of the [OIII] lines with two  (b) and three
(c) Gaussian components. Dots represent the observation and the solid line
represent the best-fitting. Gaussians are shown below.}\label{fig_04}
\end{figure*}

  There are  two  main possibilities, both triggered by a close interaction:
(i) non-circular  gas motions (first of all inwards gas streaming, see
\citet{Iono04} and references therein) and/or (ii) quasi-circular gas motions
with the orientation parameters which are different from the inner galactic
disc ones (warped outer disc, see \citet{Binney} for review).

In case (i) we can estimate the peculiar velocities via the map of residual
velocities (the differences between the observed line-of-sight velocity field
and the circular model, see Section~\ref{noncirc}). The `blue' excess of the
observed velocities ($-(30-60)$ km\,s$^{-1}$) appear near the major axis at
$r>25$ arcsec on both sides of the centre if we compare the observed data with
the model calculated for the whole disc (the full circles in Fig.~
\ref{fig_02}). On the SE side of the disc the peculiar velocities located in
the outer part of the spiral arm are directed towards the companion HCG96c. An
asymmetric shape of this part of the spiral structure is due to the
interaction with the satellites \citep{Ver97}. The line-of-sight projection of
the radial motions is zero towards the major axis; using equation (\ref{eq1})
from Section \ref{noncirc} we can obtain the residual velocities in the SE
spiral arm. That correspond to the vertical velocities of up to 70
km\,s$^{-1}$ or to the azimuthal streaming in the galactic plane of $80-120$
km\,s$^{-1}$ (in the direction of the rotation). These strong non-circular
velocities  are similar to the ones observed in the  grand design spiral
interaction galaxies, as e.g. in M51, with velocities $\pm(70-130) $
km\,s$^{-1}$ reported by \citet{Vog93}.

On the other hand, the situation on the NW side of the disc is different. The
residual velocities in the outer regions of the NW disc side have  similar
amplitude  as on the SE side, but it is only in the interarm region and it
has no connection with the  morphological structure of the disc. At the same
time the deviations from the circular model are around zero in the NW spiral
arm and in the tidal tail at the North. Moreover, if we calculate the model of
rotation only for the NW side of the disc (the diamonds in Fig.~ \ref{fig_02})
the peculiar velocities in the interarm region decrease significantly.
Therefore,  the non-circular velocities here seem to be linked to the global
distinctions in the orientation parameters of the gaseous disc at a different
values of $r$ (i.e. a warped disc).

If there is a case of warped disc,  one can  estimate the warping angle of the
disc assuming a constant rotation velocity along the radius. We found that for
$r=35$ arcsec the inclinations are $i'=(20\pm 4)^\circ$ and $i'=(48\pm
5)^\circ$ for SE and NW sides of the disc, respectively. Therefore, if the
disc is warped between $r=20$ arcsec and $35$ arcsec is moderate,
$|i'-i_0|=(13\pm6)^\circ$ for SE and  $(15\pm7)^\circ$  for  NW sides. As we
already mentioned above, it is possible for both mechanisms to contribute to
the observed kinematics: a significant non-circular motions along the SE arm
and warped disc in the NW side of the galaxy.

The radial variations in $PA_{dyn}$ around the line-of-nodes is $10-20^\circ$
in amplitude (see Fig. \ref{fig_02}). It indicates non-circular gas motions.
Exactly, at $r<7$ arcsec the dynamical axis turns in the opposite side to the
major axis of the near-infrared isophote of the circumnuclear bar with
$PA_{photometry}\approx280^\circ-290^\circ$ (see Fig.~\ref{fig_01} in
\citealt{hunt}). This type of $PA_{dyn}$ variation corresponds to the radial
streaming motion along the bar (see discussion and
 references in \citealt{mois04}). We note here that  these
motions should be bi-symmetrical, since measurements of $PA_{dyn}$
show a good agreement between the SE and NW sides of the disc. Comparise to the
bar's region, the $PA_{dyn}$ deviations at
 large distances ($r=13-22$ arcsec) are  significantly different at the SE and NW
sides of the disc. The possible explanation to this mismatch, is an influence
of the spiral arms on  gaseous disc kinematics. The shape of the spiral
pattern is  asymmetric \citet{Ver97}, hence the Northern and Southern arms
cause different amplitudes (and characters) of velocity perturbations, as it
is shown in Fig. \ref{fig_02}.

\subsection{The map of non-circular motions}\label{noncirc}

For a precise  analysis of the gas kinematics in the circumnuclear region, we
study a spatial distribution of possible non-circular motions. With this goal
we made a map of residuals between the observed velocity field in the
H$\alpha$ and the model of the pure circular rotation. The circular model was
obtained using the `tilted-rings' approximation  (see previous section) with
fixed angles $PA=PA_0$ and $i=i_0$. Fig. \ref{fig_03}a  shows a $10\times10$
kpc region of the  residual velocity field. This figure reveals a tight
correlation between large amplitude of the residuals and spiral arms traced in
the H$\alpha$-isophote. Moreover, large non-circular velocities are detected
in the
 centre ($r=1-5$ arcsec) of the galaxy along its minor axis:
$V_{res}=+(20-25) $ km\,s$^{-1}$ to NE and $V_{res}=-(40-45)$
km\,s$^{-1}$ to SW from the nucleus. Before interpreting these
features, let us recall a formula for the projection of velocity
vector in the galactic disc after the subtraction of the circular
rotation:
\begin{equation}
V_{res}= V_{r}\sin\varphi\sin i+V_{\varphi}\cos\varphi\sin i +V_{z}\cos i,
\label{eq1}
\end{equation}
where $\{V_{r},V_{\varphi},V_{z}\}$ are radial, azimuthal and
vertical component of the velocity vector respectively; $\varphi$ is the
azimuthal angle measured from the projected major axis in the plane of the
galaxy. Assuming that the motion is
only in a galactic plane, than according the first approximation, we have that  $V_z\approx 0$.
Therefore, in the points which are close to the minor
axis ($\varphi\approx90,270^\circ$) we  observe only a projection of radial
motions: $V_{res}\approx V_{r}\sin i$. The direction of these motions (inflow
or outflow) is determined by the orientation of the disc relative to the
observer. For Mrk 533 the  SW side is closer to the observer \citep{kinney},
that is in   accordance with the assumption of trailing spiral arms. In this
case, large residual velocities along the minor axis correspond to the outflow
motion $40-90$ km\,s$^{-1}$ in an amplitude\footnote{This speed should
decrease if the direction of the outflow lies out of the galactic plane, i.e
$V_z\neq0$}.

\begin{figure}
\centerline{\includegraphics[width=4.4 cm]{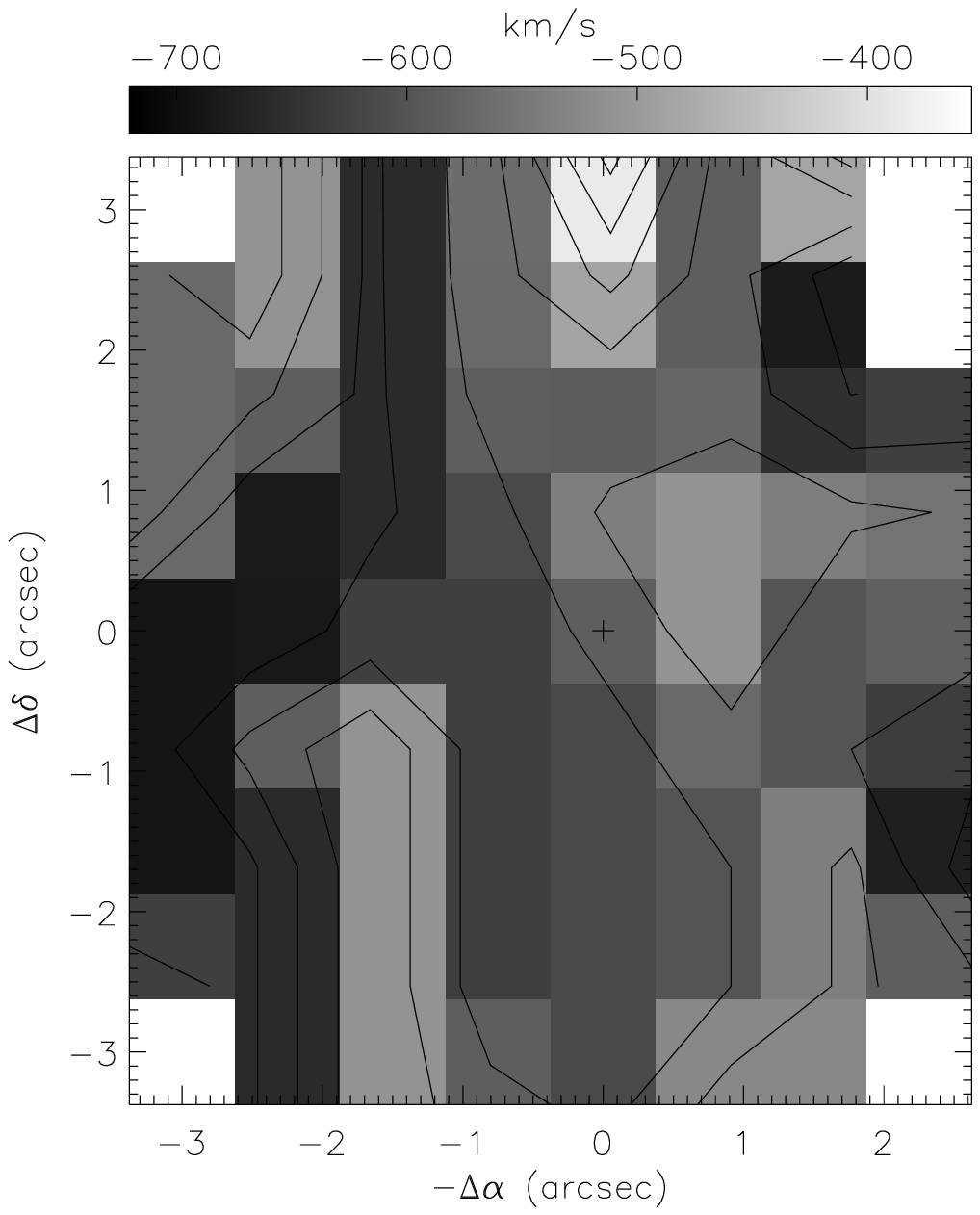}
\includegraphics[width=4.4cm]{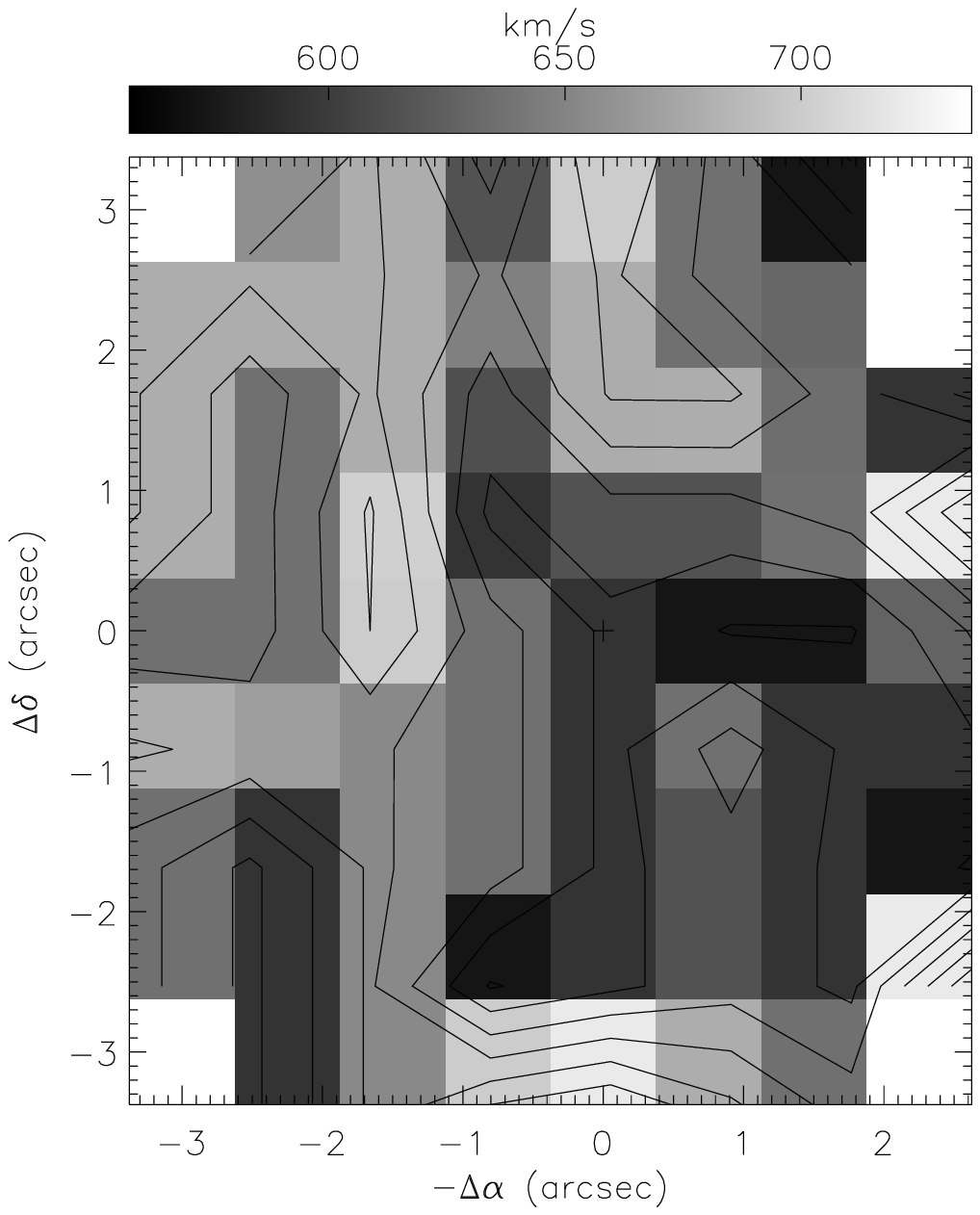}} \caption{The outflow velocity map
(left) and the field of  velocity dispersion (right) of the NLR1, where
assumption of three emission regions was applied.}\label{fig_06}
\end{figure}

Using this formalism we created a map of inflow/outflow radial streaming
motions (Fig. \ref{fig_03}b). On this map we avoid points $\mid\varphi\mid <
45^\circ$ that form with the major axis. Here the projection of the azimuthal
component of residual velocities can be larger than radial ones (for this
points should be $\mid\sin\varphi\mid<\mid\cos\varphi\mid$ because in spiral
galaxies $V_{r}$ and $V_{\varphi}$ have same amplitude, see direct
measurements by  Fridman et al. \citep{fridman1,fridman2}. Also we show in
Fig. \ref{fig_03}b the position of the bar (with orientation and axis ratio is
in accordance with the NIR photometry reported by \citealt{hunt}) and spiral
arms (see \citet{Ver97}, their Fig.~\ref{fig_06}b). The map of radial motions
shows complex structures: the outflow in the centre ($r<5$ arcsec), the gas
inflow along the Northern arm at $r=5-15$ arcsec and an outflow again at $r>15$
arcsec (8.5 kpc) along the Southern spiral arm. The difference in radial
motions for the different arms can be a result of their asymmetrical shapes
caused by an interaction with other members of the compact group of galaxies.
Also the possible position of a co-rotation at $r\approx 15 $ arcsec may
produce these changes of radial motions in the direction along the spiral arm.

\begin{figure}
\centerline{\includegraphics[width=4.4 cm]{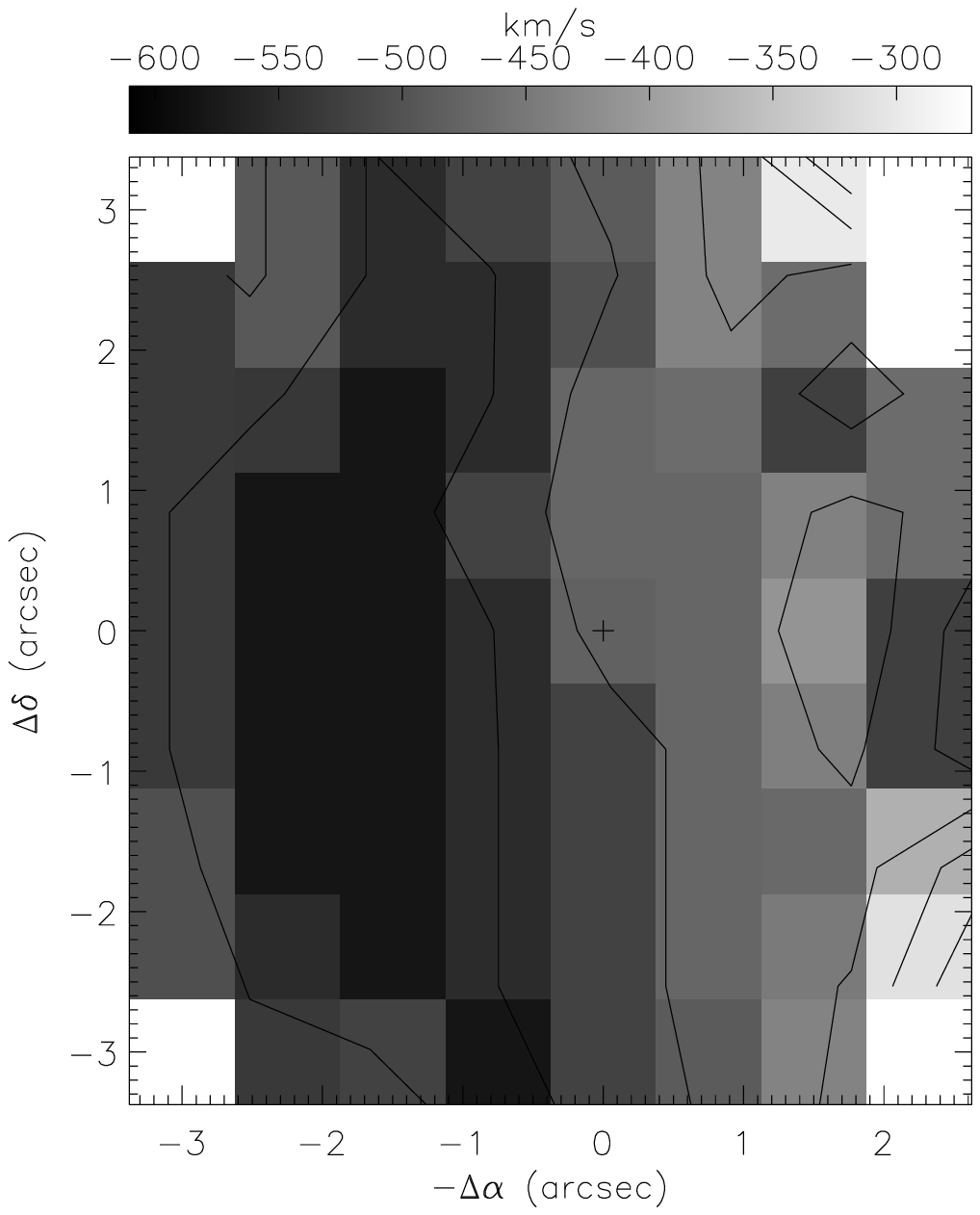}
\includegraphics[width=4.4cm]{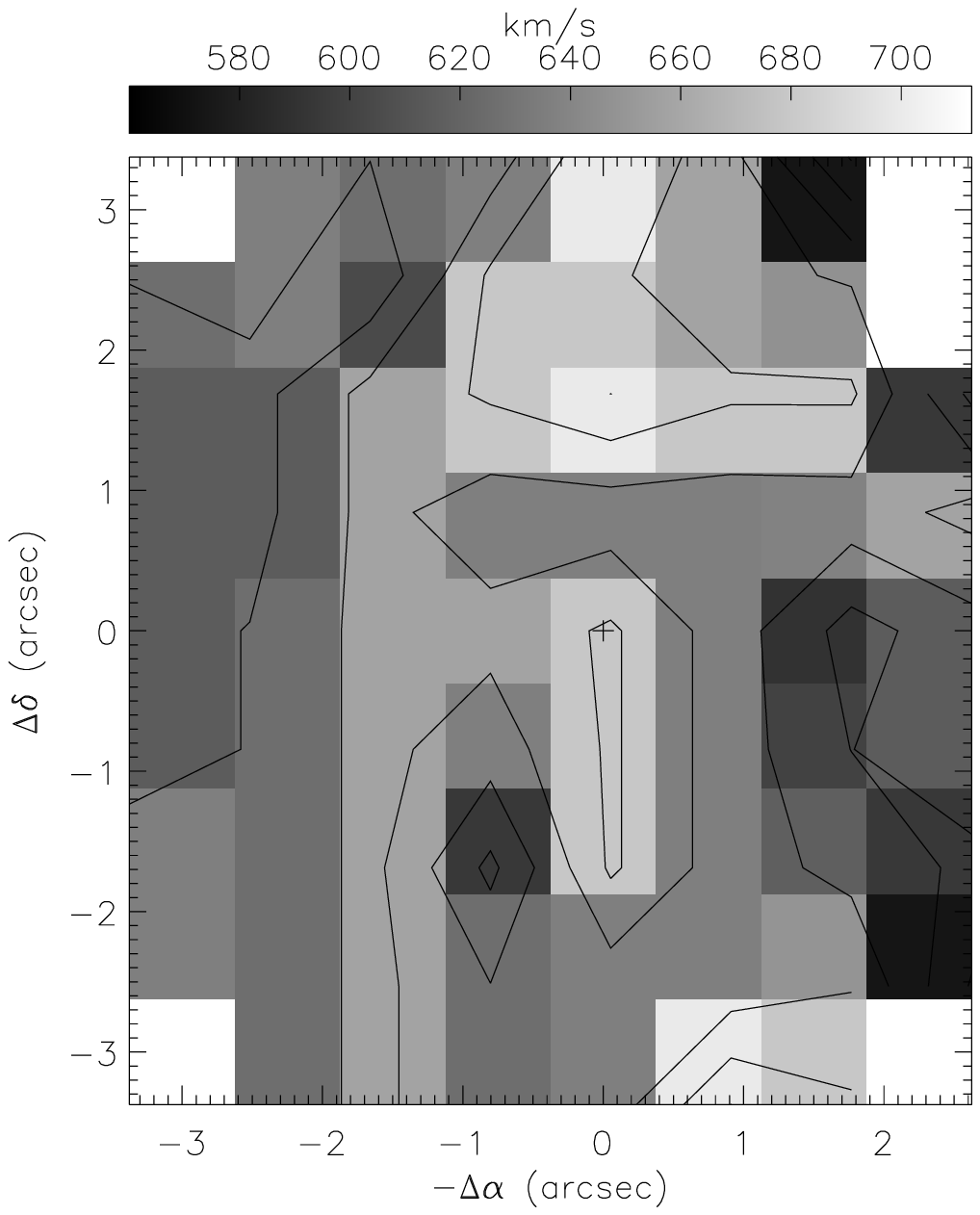}} \caption{The outflow velocity map
(left) and the field of   velocity dispersion (right) of the NLR1, where
assumptions of two emission regions were applied. The zero-point cross  marks
to the centre in the continuum images.}\label{fig_05}
\end{figure}

\section{The NLR kinematics}\label{nlr}

In  order to investigate the NLR kinematics, in this section we present the
analysis of the [OIII] emission line profiles obtained from MPFS data.  The
shapes of the narrow spectral lines have a blue asymmetry only in the nuclear
part (see Fig. \ref{fig_04}) of the galaxy. By inspecting the line profiles,
the asymmetry can be seen in the H$\alpha$, H$\beta$ and [OIII] lines, and the
shapes of the blue part of the lines differ from the various parts of the
nucleus emitting region (Fig. \ref{fig_04}). Here we have high spectral
resolution spectra only for the H$\beta$+[OIII] lines, observed in 2005, but
the H$\beta$ line is significantly weaker than the [OIII] lines and asymmetry
is not enough pronounced. Consequently we analyze in more details only the
spectra of the [OIII] lines.

It is important to note that we could spatially resolve the region with a
blue-wing asymmetry of line profiles. In order to check this, we created a map
 showing the distribution of  brightness of the flux in the broad wing of the
[OIII] lines. Indeed, our analysis showed that the FWHM of the spatial
distribution was about 2-2.4 arcsec that is significantly larger than the size
of the image of the star (seeing was 1.3 arcsec) observed with MPFS directly
before the galaxy.

\subsection{Line profile analysis}

To analyze the shape of the [OIII] lines one can fit the line profiles assuming
different functions (Lorentz, Voigt, Gaussian profiles). But taking into
account that the line profiles in the emitting region is caused by a motion of
the emitters, here we fitted each line with the sum of Gaussian components. We
applied a $\chi^2$ minimalization routine in order to obtain the best-fitting
parameters. We also assumed that the narrow emission lines can be represented
by a sum of Gaussian components. In the fitting procedure, we looked for the
minimum number of Gaussian components needed to fit the lines. To limit the
number of free parameters in the fit we set some a prior constrains
\citep{Pop04}:

(i) Two Gaussian (e.g. broad and narrow components shown in \ Fig.
\ref{fig_05}) representing the [OIII]$\lambda\lambda$4959,5007 lines  are
fixed at the same red-shift  and the Gaussian
widths are set proportional to their wavelengths,\\
\begin{center}
\begin{large}$\frac{W_{4959}}{4959}=\frac{W_{5007}}{5007}$\\\end{large}
\end{center}
(ii) We assumed that the intensity ratio of the two
[OIII]$\lambda\lambda$4959,5007 lines is $\approx$1:3 \citep{Dim07}.

Here we will use the velocity dispersion to present the velocity field
($\sigma=(W/\sqrt{2}) \cdot (c/\lambda$), where $W$ is the Gaussian width).
Outflow velocities are considered to be
$V_{r}=c\times\frac{\bigtriangleup\lambda}{\lambda}$, where
$\bigtriangleup\lambda=\lambda_{c}-\lambda_{b}$ is a blue shift, measured
between the centre of the  narrowest component that is assumed to correspond
to the transition wavelength $\lambda_{c}$ and those which are blue shifted
$\lambda_{b}$ (Fig. \ref{fig_04})

\begin{figure}
\begin{center}
\centerline{\includegraphics[width=4.4 cm]{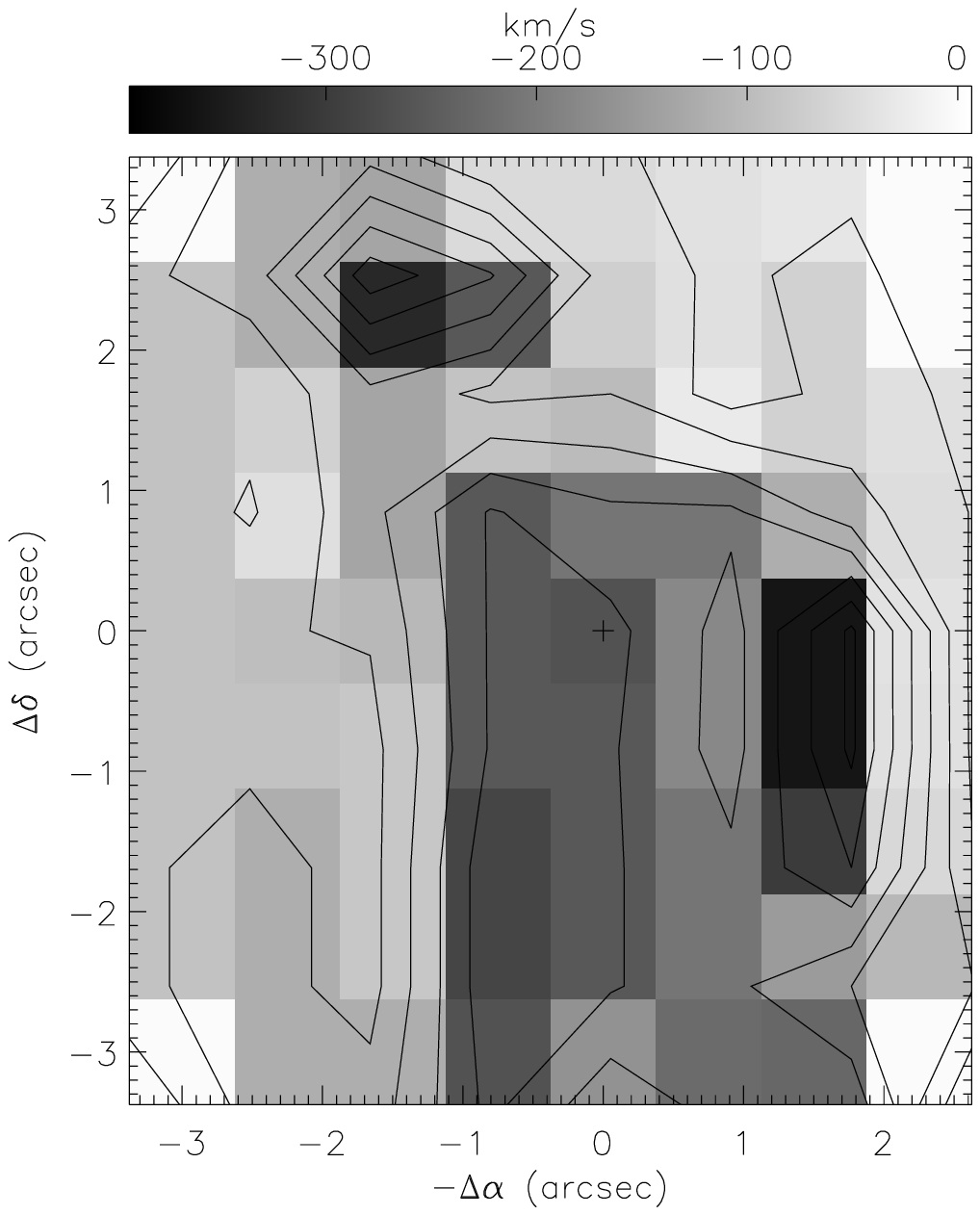}
\includegraphics[width=4.4 cm]{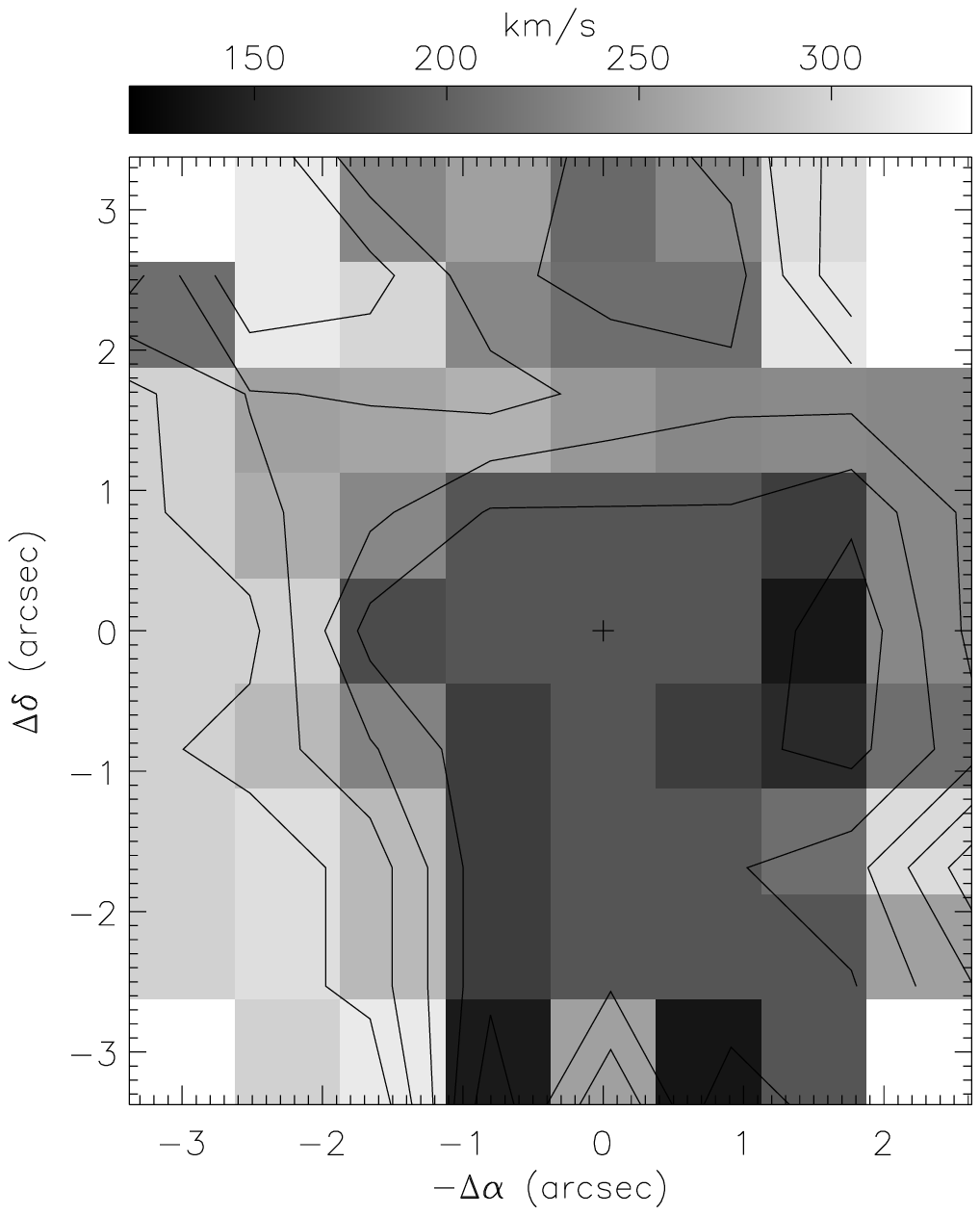}}
\caption{The outflow velocity map (left) and the field of  velocity dispersion
(right) of the NLR1a.}\label{fig_07}
\end{center}
\end{figure}

\begin{figure}
\begin{center}
\centerline{\includegraphics[width=4.4 cm]{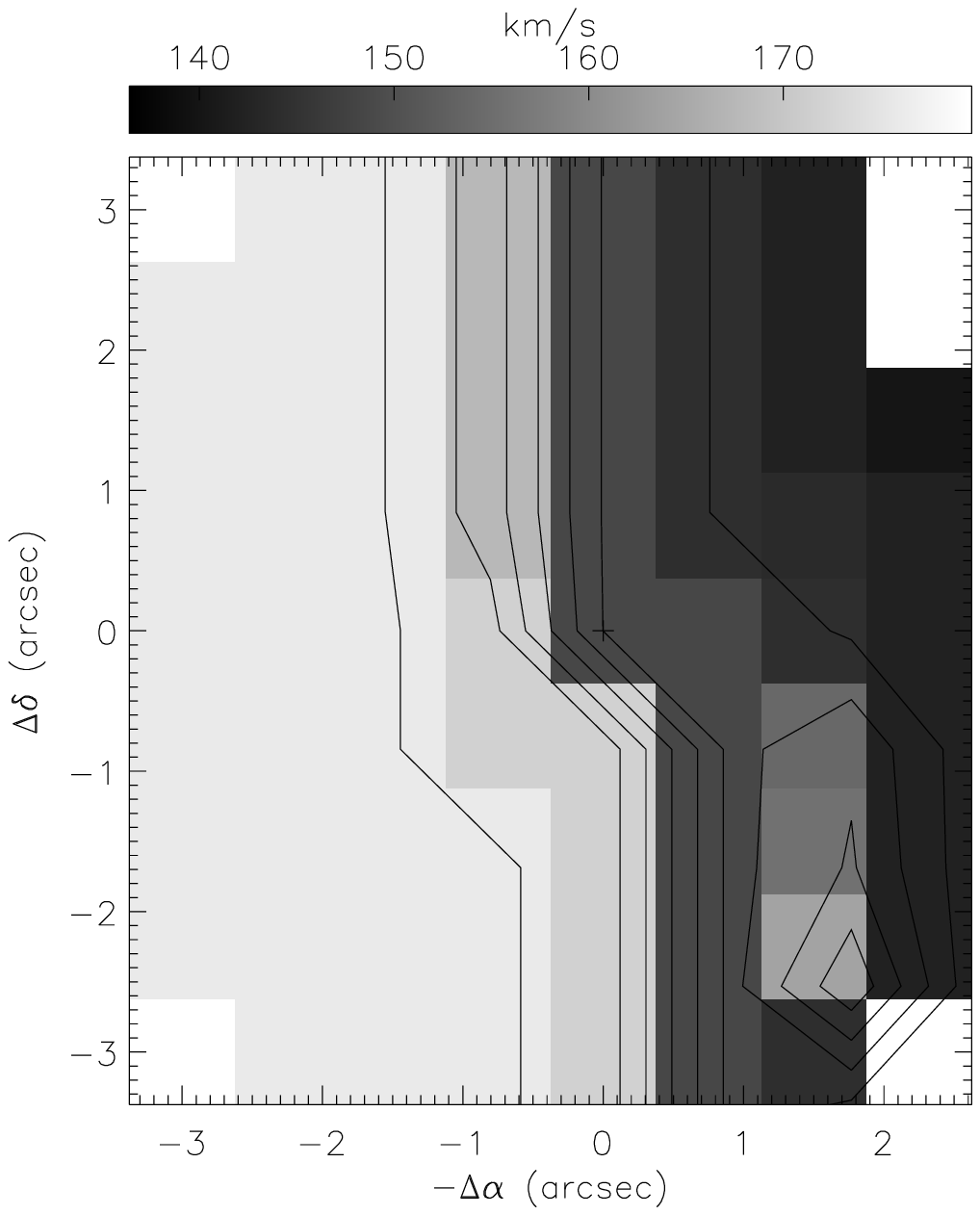}
\includegraphics[width=4.4 cm]{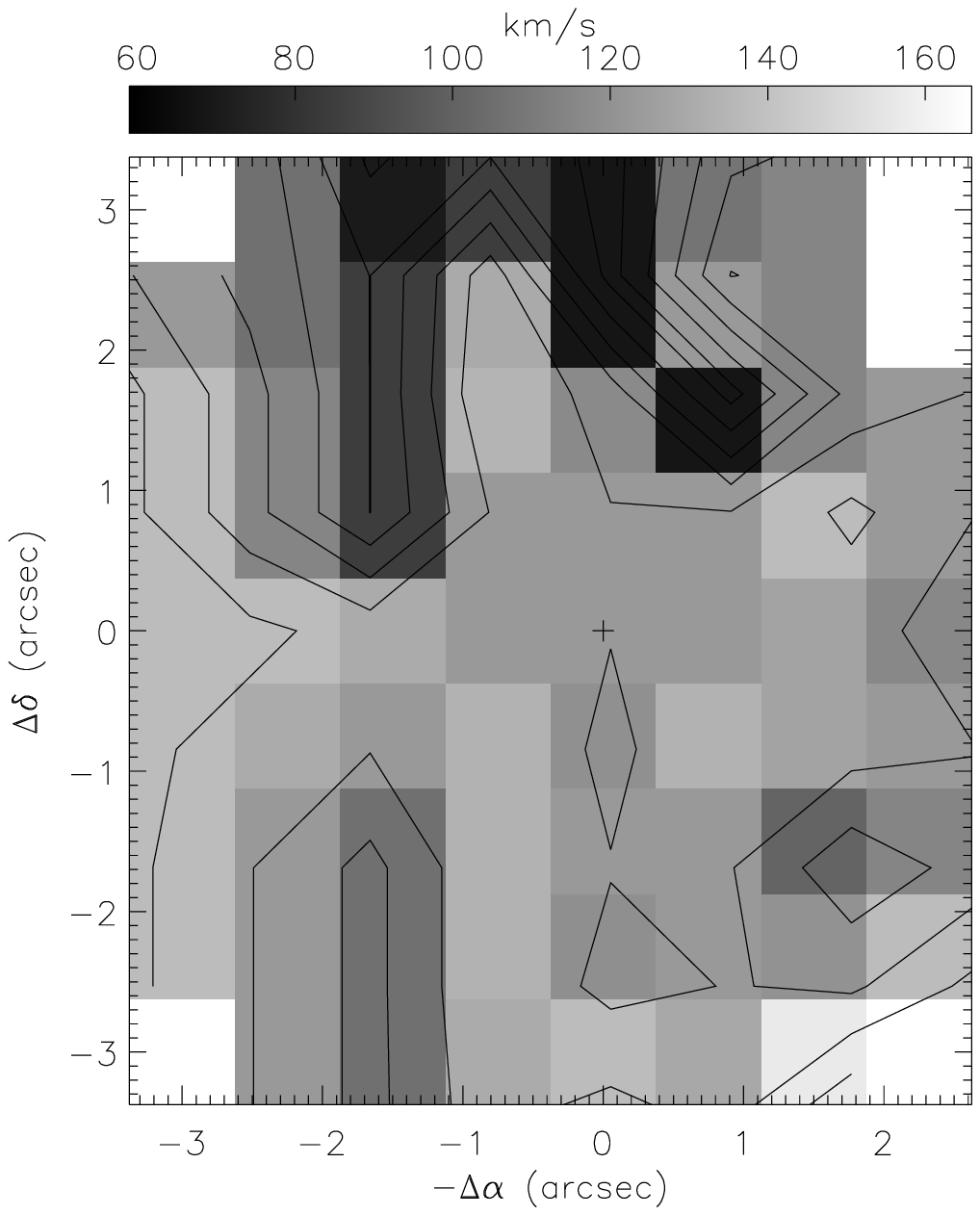}}
\end{center}
\caption{The field of  velocity dispersion of the NLR2: 1. The case of fitting
spectra with two Gaussians (left); 2. For the case of fitting spectra with
three Gaussians (right).}\label{fig_08}
\end{figure}

\begin{figure}
\begin{center}
\centerline{\includegraphics[width=4.4 cm]{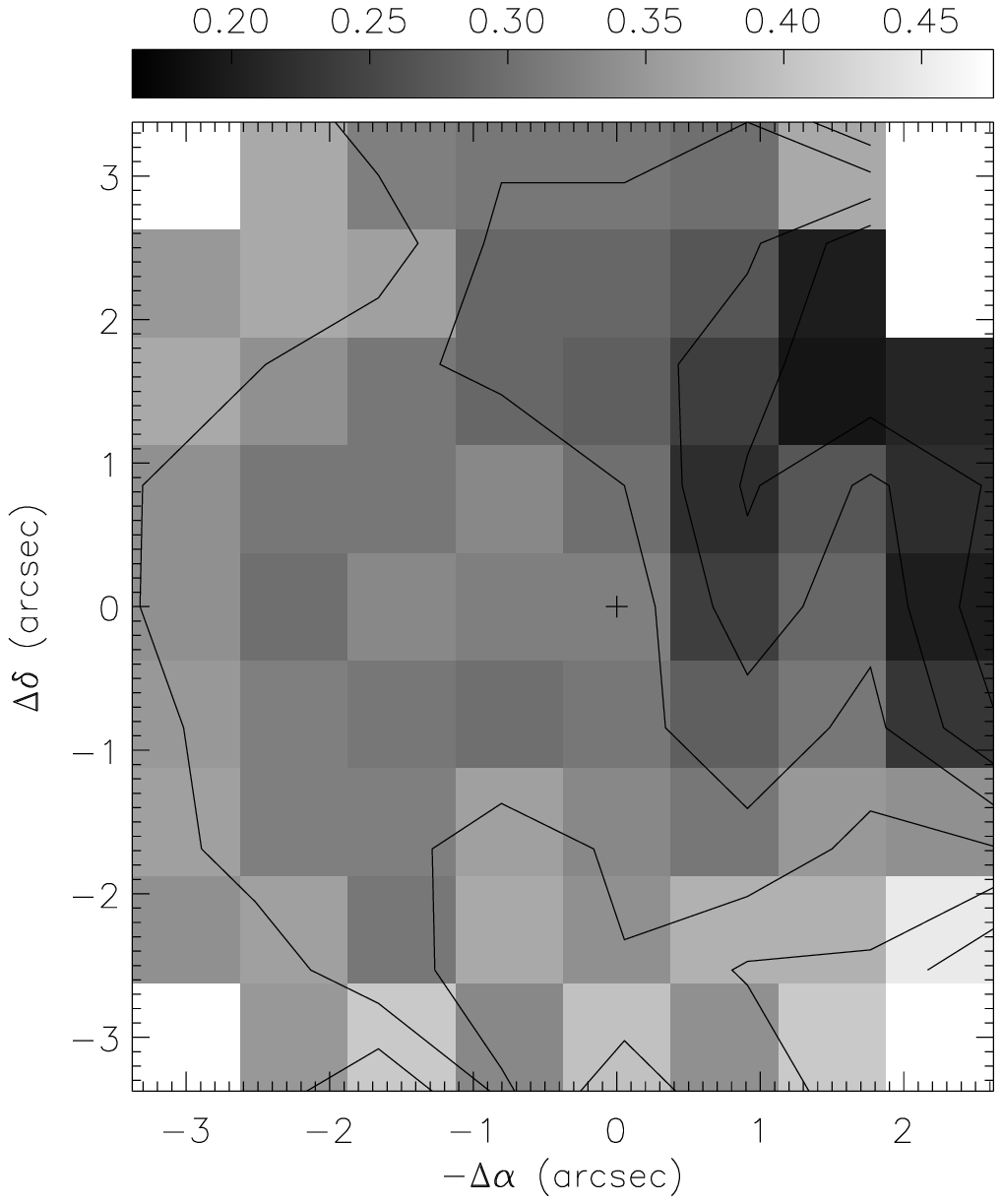}
\includegraphics[width=4.4 cm]{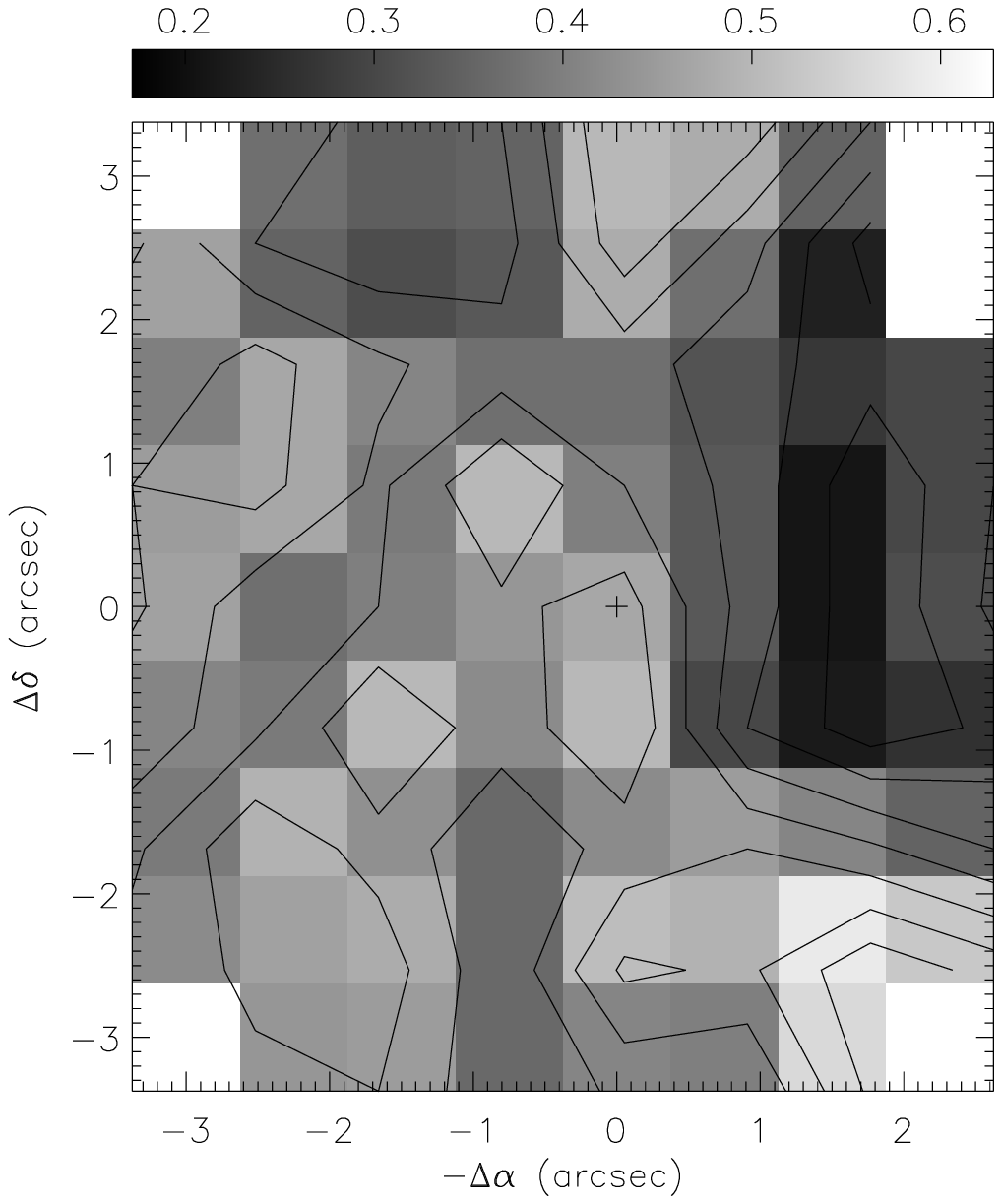}} \caption{Flux ratio
between NLR1 and NLR2 emission components where two (left) and three (right)
emission regions are assumed.}\label{fig_09}
\end{center}
\end{figure}

As mentioned above (see in Fig. \ref{fig_04}) the [OIII] lines show a blue
asymmetry, and can be properly fitted with two Gaussian components (Fig.
\ref{fig_04}b). Moreover, a slightly better fit (concerning $\chi^2$) can be
obtained by fitting with three Gaussians (Fig \ref{fig_04}c). Therefore, here
we will consider the complex [OIII] line region as composed from two or three
separated emission regions.
 When we assume the emission of two regions,
we denote those regions as the NLR1 (blue shifted component) and NLR2
(non-shifted component). In the case where we assume that the emission is
coming from three regions, we additionally consider a NLR1a existence (also
the blue-shifted component) region that corresponds to a  component placed
between the central and broad blue shifted one (Fig. \ref{fig_04}c). We started
to fit each spectra obtained from spatial MPFS elements, but out of the
nucleus (around 8$\times$9 elements) it is not possible to see the asymmetry,
so the method of fitting analysis has been applied only in the small frame
elements that corresponds to a small region on the sky around $6 \times 7$
arcsec.

The velocity maps that we obtained from our analysis are presented in Fig.
\ref{fig_05}-\ref{fig_08}. Also we found a map of flux ratios of NLR1 and
NLR2, i.e. $\frac{F_{NLR1}}{F_{NLR2}}$ (Fig. \ref{fig_09}).

\subsection{Assumption of two emitting regions}

Assuming the presence of two kinematically separated regions, we fitted the
[OIII] lines with two Gaussians (Fig. \ref{fig_04}b). As one can see in Fig.
\ref{fig_04}b the line shapes can be well fitted with two Gaussians, but a
small shoulder in the blue part can not be fitted properly with this
assumption. The map of the outflow in the NLR1 is presented in Fig.
\ref{fig_05} left. As can be seen in this figure, the maximum outflow velocity
($\sim -550 $ km\,s$^{-1}$) is located to the east from the continuum centre
(presented as a cross in Fig. \ref{fig_05}). A gradient in outflow velocities
from the east to the west is present. The obtained difference amongst
velocities (from the east to the west) is $\sim 300 $ km\,s$^{-1}$. This
supports the idea about the existence {of} an {approaching} jet from the
active galactic nucleus \citep{Mom03}. The velocity field of the outflow has a
south-north elongated structure (Fig. \ref{fig_05}a), which might be due to
the projection of the jet. On the other hand the  velocity dispersion field
has a relatively small difference in NLR1 ($\sim560$-670 km\,s$^{-1}$),
showing a relatively flat velocity field. The intensity ratios in the central
part of NLR1/NLR2 are present in Fig. \ref{fig_09} (left). The maximal
intensity of the NLR1 is around the optical centre.

The NLR2  velocity dispersion map shows also stratification, where
 velocity dispersion has values ranging from 150 km\,s$^{-1}$ to 170 km\,s$^{-1}$,
having a gradient from the west to the east (Fig. \ref{fig_08} left). This
gradient in  velocity dispersion has a possible connection with the radio jet
which is also elongated in the same direction (see Section \ref{rad}).
However  this should be taken with caution, since the differences in  velocity
dispersion are too small (around 20 km\,s$^{-1}$, that is very close to the
range of the error bars of the fitting method).

\subsection{Assumption of three emitting regions}

As was mentioned above, a slightly better fit can be obtained when we perform a
three Gaussian fitting procedure. Moreover,  there is an anticorrelation
between the outflow velocity map of the NLR1 and dispersion velocity map of
NLR2 (see Figs. \ref{fig_05} and \ref{fig_08}) that may be caused by fitting
the observed profile with an insufficient number of components (only two
components). Therefore, we assume that in the nucleus of Mrk 533 three,
kinematically separated, emission regions are present on the line-of-sight.
The velocity maps of the NLR1 and NLR1a are presented in Figs. \ref{fig_06}
and \ref{fig_07}. As one can see in Fig. \ref{fig_06} the outflow and velocity
dispersion in NLR1 are slightly higher then in the case of two region
approximation. The maximum outflow velocities ($\sim -700$ km\,s$^{-1}$) are
located to the east of the optical centre, having a south to north elongated
structure. In comparison to the two region approximation, the three region
approximation shows that the outflow structure is more complex then the one
obtained from two Gaussian fit. Also, the map of  velocity dispersion shows
several regions around the optical centre with higher velocities (Fig.
\ref{fig_06}, right). But this should be taken with caution, since the
difference in  velocity dispersion is small ($\sim 100 $ km\,s$^{-1}$) and the
 velocity dispersion field of the nucleus of Mrk 533 tends to have a flat
distribution, as it was the case in the two Gaussian assumption.

The NLR1a region shows a complex structure around the optical centre, with
maximum outflow velocity around  -250 to -300 km\,s$^{-1}$, and minimum
velocity dispersion around 140 -- 320 km\,s$^{-1}$, see Fig. \ref{fig_07}.
Also, the ratio between NLR1 and NLR2 fluxes (Fig. \ref{fig_09}) indicates a
clumpy NLR.

In this case the NLR2 velocity dispersion region is similar to the one in two
Gaussian case, where  velocity dispersion has values ranging from 70
km\,s$^{-1}$ to 160 km\,s$^{-1}$ (Fig \ref{fig_08}). As one can see there are
some structures which might indicate also a clumpy NLR2.

\subsection{The complex NLR: Two or three emitting regions?}\label{cnlr}

 Based on our analysis the [OIII] line shapes across the central part of
Mrk 533, it can easily be noted that there is  an outflow. The outflow
 also can have a complex structure  in velocity and
spatial domain. Although the [OIII] line profiles can be well fitted with the
assumption of two emitting region (where one is characterized by the presence
of an outflow), the better fit of the line shapes can be reached with the
three emitting region assumption. Moreover, the anticorrelation between the
NLR1 outflow velocity map and the NLR2 velocity dispersion map in the two
emission region indicates  that there are probably two regions with outflows.
Here the outflow velocities range from -500 to -700 km\,s$^{-1}$ in one region
and range from -200 to -400 km\,s$^{-1}$ in the another. The obtained
velocities are in  accordance with the outflows registered from the O VI
absorption lines,  where the measured velocity of the outflows are -800 and
-300 km\,s$^{-1}$ \citep{Shas04}. It seems that these regions in outflow also
absorb in the UV radiation. On the other hand, our research of the
circumnuclear region (see Section~\ref{kin}) shows that near the nucleus $r<5$
arcsec there is an outflow, but with a smaller outflow velocity (from -40 to
-90 km\,s$^{-1}$). This type of outflow was observed only in the narrow
component of the H$\alpha$ profile when  the circular rotation velocities were
taken into account.

\section{Connection between  the radio, UV and optical outflow}\label{rad}

In this part we compare our results obtained in the optical range with the
previous results which were obtained in the UV  and radio band in order to
connect the outflow(s) seen in the NLR1 with the radio jets. An indication of
a jet-cloud interaction is the ratio of the emission lines (which can
originate from shocks), therefore we first explore the nature of the gas
ionization source in the nuclear and circumnuclear region as well as in the
spiral structure.

\subsection{The source of ionization}

\label{ion}

\begin{figure*}
\includegraphics[width=8.5cm]{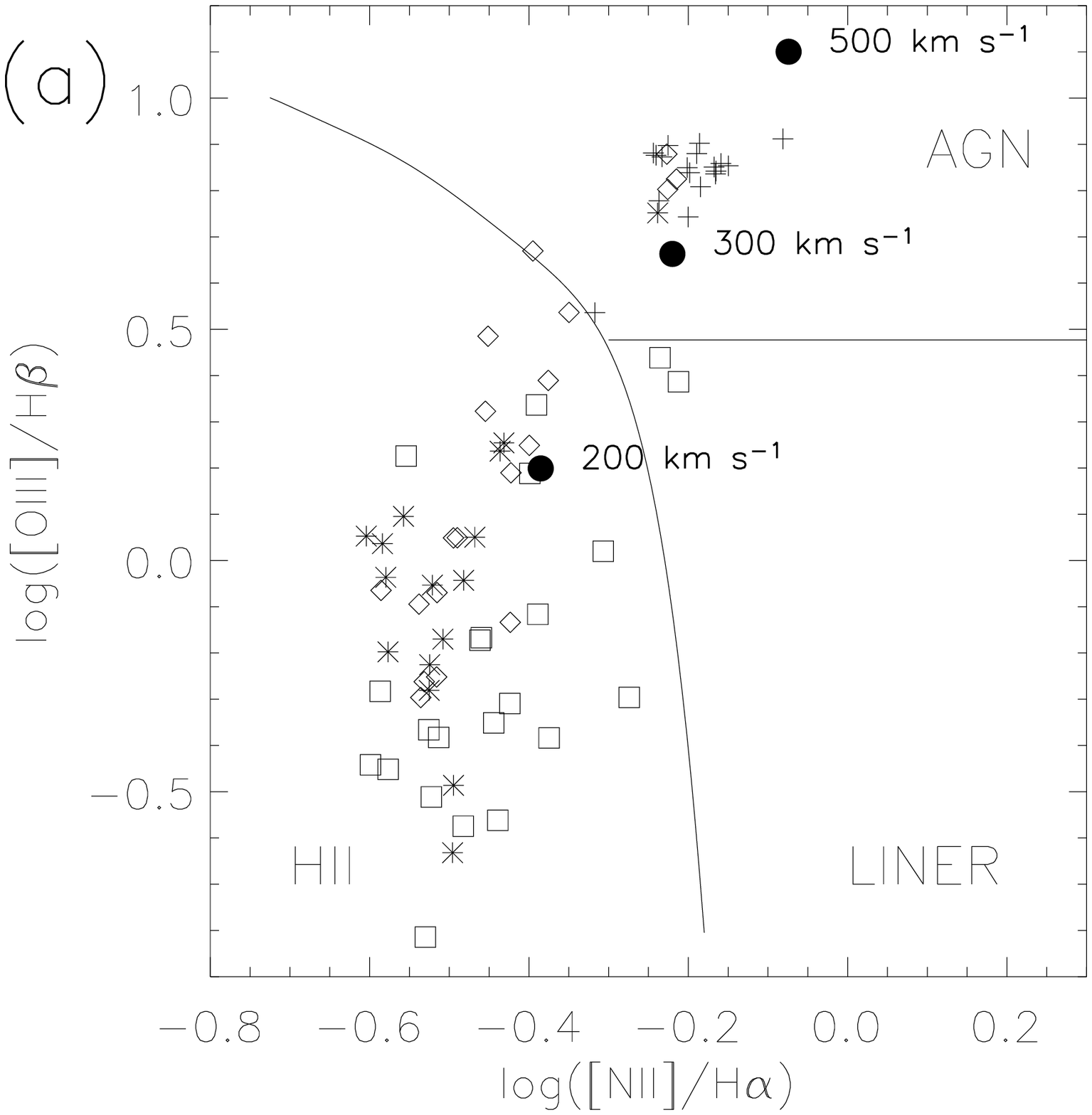}
\includegraphics[width=8.5cm]{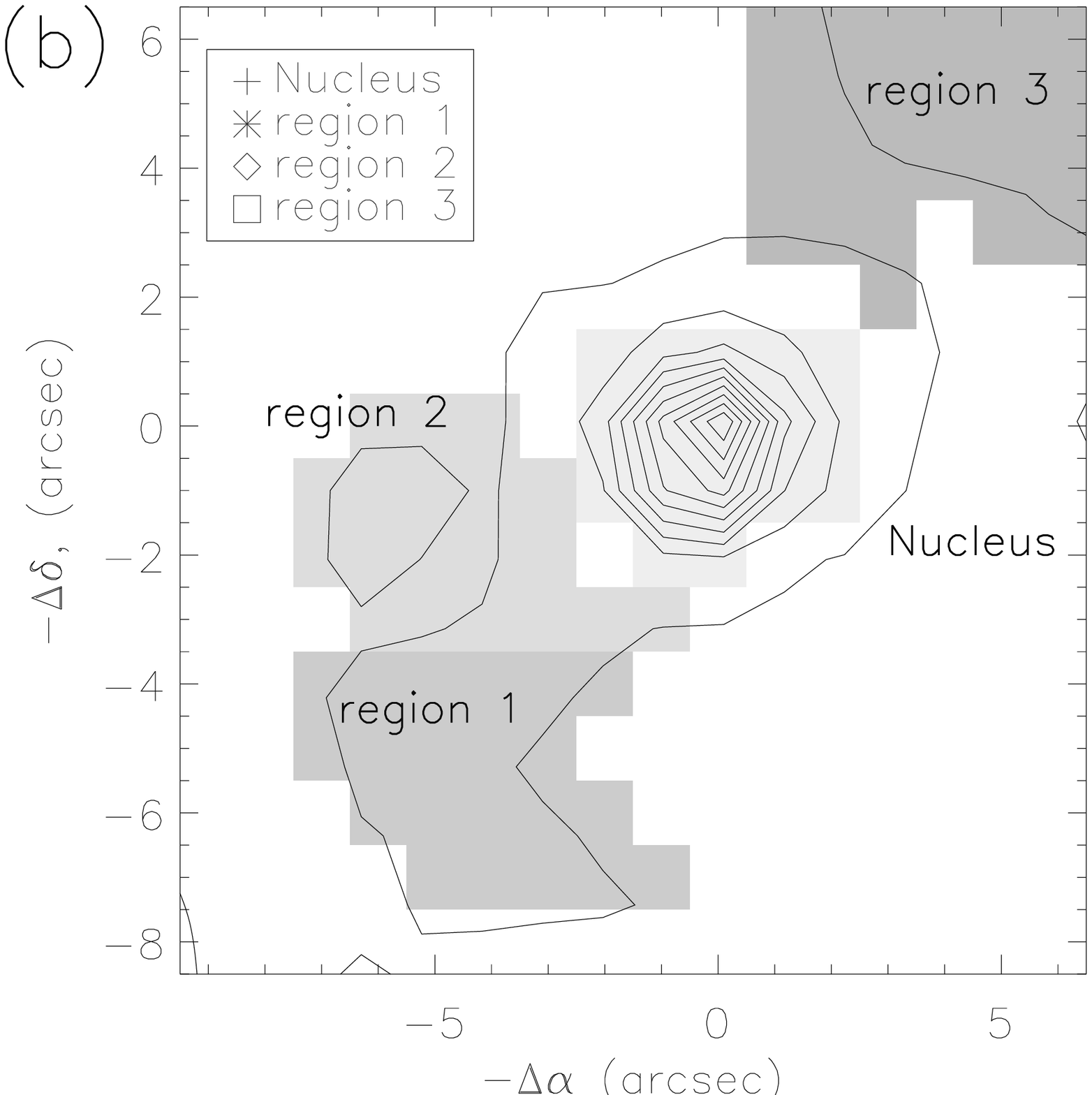}
 \caption{The flux ratio of [OIII]/H$\beta$ lines vs. [NII]/H$\alpha$
(a) and the map of Mrk 533 with regions for which this ratios are found (b).
The H$\alpha$ isophotes are overlapped on this map. The full circles  in Fig.
4a represent the calculated ratio by \citet{Dop95}, where 'shock plus
precursor' model were used. The calculated values are for the shock velocities
of 200 km s$^{-1}$, 300 km s$^{-1}$, and 500 km s$^{-1}$ (the values are given
near the full circles).}\label{fig_10n} \end{figure*}

 In order to identify the nature of the gas ionization source in
Mrk 533 we constructed diagnostic diagrams using the emission line intensity
ratios, i.e. in Fig. \ref{fig_10n} we present [OIII]/H$\beta$ vs.
[NII]/H$\alpha$. According to \citet{Vei89} we separated the regions
corresponding to the ionization by AGN, young OB-stars (HII regions) and shock
waves (LINERS) into presented diagram. As a result of short analysis of the
spectra, we were able to separate several emitting regions (see Fig.
\ref{fig_10n}, right). Mainly there are three emission regions in the stellar
disc plus an emitting region that corresponds to the nucleus. The analysis of
the flux ratio of [OIII]/H$\beta$ vs. [NII]/H$\alpha$ lines shows that spectra
belonging to the nucleus (crosses) are located in the part where the
ionization of AGN is expected. On the other hand, spectra from the spiral-like
emission structure (regions 1, 2 and 3) are mainly located in the HII part of
the diagram, i.e. the ionization from hot young stars is present. Note that
some points from the region 2, that is closest to the AGN, are located also in
the AGN part of the diagram. It means that  non-thermal ionization
predominates in some parts of the region 2. It is interesting that some points
from region 3 are located near the HII/LINER borderline. This may indicate
that besides the star bursts ionization, gas may be compressed and that in
parts of this region, a shock ionization occurs.

From the inspection of Fig. \ref{fig_10n} it is evident that  different
ionization mechanisms may dominate on different scales. We find that in nucleus
part as well as in the regions (1 and 2) which are close to the nucleus the
dominant ionization mechanism is AGN-photoionization. However, as we move
further away from the nucleus ($\sim$ kpc), jet-cloud interactions may become
the main ionization mechanism. In order to discuss the possibility of the
dominance of a jet-cloud interaction in some emission regions of Mrk 533, let
us recall results obtained by \citet{Dop95}. They investigated the line fluxes
expected from fast, radiative shocks propagating through typical narrow-line
AGN clouds. Here, the key parameters controlling the emission-line ratios are
shock velocity and the magnetic parameter. \citet{Dop95} found that inclusion
of 'precursor' ionization caused by hard radiation generated in the wake of
the post-shock, cooling plasma that diffuses outwards, is crucial for
explaining differences found in Seyfert and LINER galaxies. It is interesting
that faster shocks are required ($\sim$ 300 -- 500 km s$^{-1}$), that is well
fitting our obtained outflow velocities in the NLR1 and NLR1a. We present in
Fig. \ref{fig_10n} with full circles the predicted line ratios for different
shock velocities ('shock plus precursor', see \citet{Dop95}).  The observed
[OIII]/H$\beta$ vs. [NII]/H$\alpha$ ratios in the nucleus region are
consistent with a 'shock plus precursor' model that has shock velocity
exceeding ~300 km s$^{-1}$ at least, and a magnetic parameter typical for the
interstellar medium \citep{Dop95}. Moreover, some points from the other three
emission regions shown in Fig. \ref{fig_10n}b are also consistent with a
'shock plus precursor' model, but with smaller velocities $\sim$ 200 km
s$^{-1}$.  Note here that the flux ratios of [OIII]/H$\beta$ and
[NII]/H$\alpha$ are constructed only using narrow lines observed with the MPFS.

\subsection{Outflows in the radio, optical and UV emission}

\begin{figure}
\begin{center}
{\includegraphics[width=8.5cm]{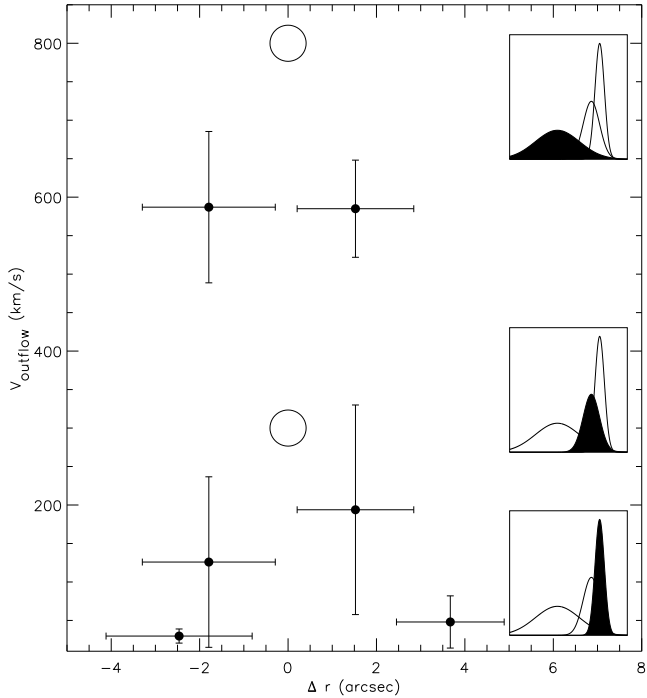}}
 \caption{The outflow velocities in the central part of Mrk 533 (within $-4.5^{\prime\prime}
<r<5^{\prime\prime}$) obtained from  Gaussian fits of the [OIII]
lines (nuclear region) and from the analysis of circumnuclear
kinematics of H$\alpha$ (full circles, crosses indicate the
interval of scattering $V_{outflow}$ and r). The circles represent
the outflow velocities from absorption components in O VI lines
\citep{Shas04}. The diagrams on the right show a component of the
emission profile (marked as black Gaussian) corresponded to the
different scales of $V_{outflow}$.}\label{fig_11}
\end{center}
\end{figure}

\begin{figure}
\begin{center}
{\includegraphics[width=8.5cm]{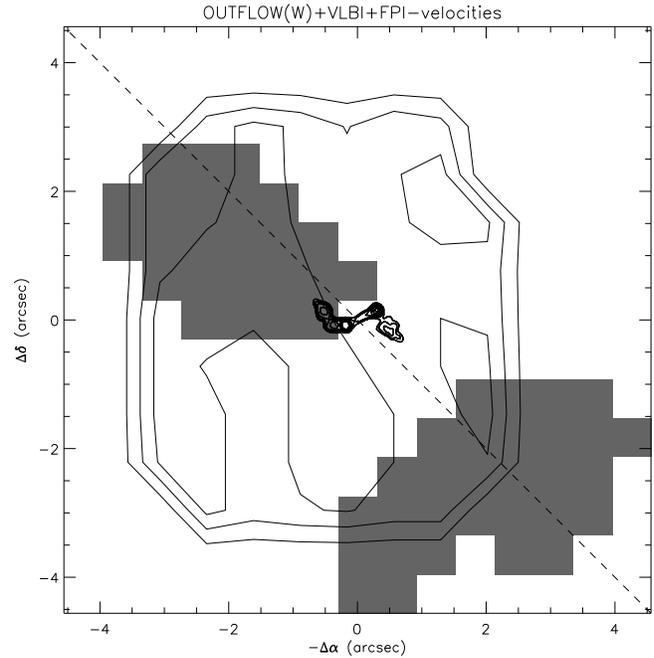}} \caption{The optical
vs. radio outflow: Gray colors shows the position of circumnuclear
outflow region seen on Fig. \ref{fig_03}b. The isolines show the
outflow mapped using the [OIII] lines (Fig. \ref{fig_06}, left)
and feature in the centre is the map of the radio continuum given
by \citep{Mom03}.  The dashed line marks the direction of the
x-axis on  Fig. \ref{fig_11}}\label{fig_12}
\end{center}
\end{figure}

To reconstruct the picture obtained from the analysis of the NLR and
circumnuclear kinematics we draw Fig. \ref{fig_11}, where the outflow
velocities in all three components are given. The velocities are measured
along the dashed line shown in Fig. \ref{fig_12}.  The circles in Fig.
\ref{fig_11} represent the measured velocity of the outflows  from O VI
absorption lines (-800 and -300 km\,s$^{-1}$, see \citealt{Shas04}). As it can
be seen in Fig. \ref{fig_11}, there is a stratification in the line-of-sight
velocities, that may also indicate a different distribution of these
components, as well as, heaving in mind the different absorption densities of
the UV. The spatial structure of the outflow is slightly resolved: the
high-velocities of ($V_{outflow}>100$ km\,s$^{-1}$) NLR1 extends up to 1.5
kpc, and the low-velocity ($V_{outflow}=-(20-80)$ km\,s$^{-1}$) emission
regions extend up 2.5 kpc (see Fig. \ref{fig_11}). We remind that our spatial
resolutions were about 0.7 kpc (1.3 arcsec) on MPFS and 1.4 kpc (2.5 arcsec)
with FPI. Since the emission originates far away from the nucleus, one can
consider the possibility that the mechanism for the optical line emission may
actually be a shock wave, created in a wind (or jet) from the AGN,  that is
also in accordance with the analysis of the source of ionization (see
Section~\ref{noncirc}).

Note here, that with the MPFS and FPI we have detected essentially different
components of the outflow. In the first case the multicomponent fitting of the
emission lines profiles  was carried out and the high-velocity component in
the blue wing of the emission lines profiles was studied. In the second case,
the narrow spectral range of the FPI did not allow us to analyze the shape of
the spectral profile.  However, higher (in comparison with MPFS) spectral
resolution  gives the most accurate measurements of the velocities of the
emission line core. Moreover, using a large field of view and  better spatial
sampling we can  determine the differences between the observed velocities and
the circular rotation model. Therefore, we can distinguish the peculiar
velocities with amplitudes about several tens of km\,s$^{-1}$.  We illustrated
the differences between the MPFS and the FPI measurement on the right side of
Fig. \ref{fig_11} where  components of the emission lines correspond to the
given outflow velocities shown schematically.

To find any connection between the optical and the radio outflow (reported in
\citet{Unger88}, \citet{Mom03}, \citealt{midd04}) we draw a map with the
optical and radio outflow structures (see Fig. \ref{fig_12}). It is indicative
to see that the orientation of two jets from two points such as radio sources,
are directed to the optical outflow structures. Also, as one can see in Fig.
\ref{fig_12}, the maximal outflow velocities correspond to the side where the
most intensive part of the radio structure is located. Concerning the map
shown in Fig. \ref{fig_12}, one can speculate that the optical outflow is
following the radio one, i.e. that the jet emits in the radio domain, as it is
close to the central engine ($r<1$ arcsec or 0.5 kpc), and farther from the
centre of the  UV absorption/emission (see Fig. \ref{fig_11}) and optical
emission occurs ($r=2-4$ arcsec or $1-2$ kpc). We outline that this optical
emission, partly triggered by the shock waves originates from the jet
intrusion in to the gaseous ambient medium.

 The presence of high-velocity clouds, located close to the
passage of the radio jet and centred near the centre of the nucleus, shows
that the effects of jet-induced shocks are still important in the nuclear
regions. Moreover, the velocity dispersion of the  [OIII] component in the
NLR1 is high: ranging from 570 to 720 km s$^{-1}$. Increased linewidths are
expected following turbulent motions of an interaction (e.g. by jet-cloud
interaction), but not in passive photoionization by the AGN (see e.g.
\citet{Clark,Gand06}). On the other side in both approximation (two and three
emission line region) we noted outflows with a relatively high velocity.
Therefore, the assumption that the projected radio jets (see Fig.
\ref{fig_12}) have an influence on NLR1 physics and kinematics seems to be
real.


\section{Conclusion}\label{concl}

In this paper we analyzed the spectroscopic observations of
Mrk 533
 performed by the 6-m telescope of SAO RAS. We investigate the gas kinematics in the
spiral and circumnuclear structures. Based on this investigation we are able to
conclude the following:

i) We found the parameters of the galactic gaseous disc orientation:
$i_0=(33\pm 5)^\circ$ and $PA_0=(306\pm 4)^\circ$. The parameters are in
accordance with the ones obtained by \citet{Ver97}, but they are significantly
different from the ones measured by \citet{amram03}.

ii) Our analysis of the large-scale ionized gas velocity field shows presence
of significant non-circular motions. Also we indicate the outer warping
gaseous disc, probably caused by an interaction with the companions. The disc
warping was found to be around $13-15^\circ$ at distances $r>20$ kpc (at least
in the NW side of the disc).

iii) We created a  map of radial motions in the gaseous disc. It
shows a very complex non-circular gas motion in Mrk 533, that may be
triggered by an interaction. So, the gas inflow along the Northern
arm ($5''<r<15''$) and streaming motions along the bar were found
by analyzing of this map and as well as radial dependence of the
dynamical axis $PA$. A co-rotation at $r\approx8.5$ kpc may produce
changing of radial motions direction along the Northern spiral arms
at $r>15$ arcsec. The circumnuclear outflow in $r<5$ arcsec along
the minor galactic axis was also detected.

iv) The central emitting region of Mrk 533 is complex and can be
divided into: the emission of the Sy 2 nucleus (in the very
central part) and the star-forming regions which follow the spiral
structure of the galaxy. The emission line shapes in the region of
the nucleus have a blue asymmetry this indicates an outflow from
the centre.

To investigate  the velocity structures of the nucleus, we assumed
that the NLR of Mrk 533 is composed of: (a) two and (b) three
kinematically separated regions. From this analysis we can
conclude that:

-- The [OIII]$\lambda\lambda$4959,5007 lines originate from at least two
kinematically separated  regions: NLR1 contributes to the blue part of the
spectral lines, where a gaseous outflow with velocities from -300 km\,s$^{-1}$
(5 arcsec far from the centre) to -600 km\,s$^{-1}$ around the nucleus (the
velocity dispersion is around 560 -- 670 km\,s$^{-1}$) exist; NLR2 has a
smaller  velocity dispersion ($\sim$ 150 km\,s$^{-1}$).

-- The complex [OIII] lines may be produced in three kinematically separated
regions: NLR1 contributes to the blue part of the spectral lines, with a
gaseous outflow with velocities ranging from -500 to -700 km\,s$^{-1}$; the
NLR1a with outflow velocities from -350 to -100 km\,s$^{-1}$ and NLR2 without
a significant outflow.

Finally, we can conclude that a stratification in the NLR of Mrk 533 exists,
and that it has a range of outflow  velocity  from 20--50 km\,s$^{-1}$ to
600--700 km\,s$^{-1}$ on radial distances $\sim2.5$ and  $\sim1.5$ kpc
accordingly. The structure of the optical outflow seems to be complex, but it
seems that  the registered outflow(s) in the optical corresponds to the one(s)
registered in the   UV. Moreover, it is likely that the outflow obtained from
the optical emission lines ([OIII] and H$\alpha$) follows the radio one and
that the complex structure of the optical outflow is generated by the radio
jet.

\section*{Acknowledgments}

This work is based on observations carried out at the 6-m telescope of the
Special Astrophysical Observatory of the Russian Academy of Sciences, operated
under the financial support of the Science Department of Russia (registration
number 01-43). This research has made use of the NASA/IPAC Extragalactic
Database (NED) which is operated by the Jet Propulsion Laboratory, California
Institute of Technology, having contract with the National Aeronautics and
Space Administration.  This work was partly supported by the Russian
Foundation for Basic Research (project nos. 06-02-16825). AAS also thanks the
Russian Science Support Foundation. This work is a part of the project
(146002) `Astrophysical Spectroscopy of Extragalactic Objects', supported by
the Ministry of Science of Serbia. We would like to thank to the anonymous
referee for very useful suggestions which helped us improve the paper.

\end{document}